\documentclass[a4paper,11pt]{article}
\pdfoutput=1
\usepackage{jheppub,MnSymbol}

\newcommand{\beq}{\begin{equation}}
\newcommand{\eeq}{\end{equation}}
\newcommand{\bea}{\begin{eqnarray}}
\newcommand{\eea}{\end{eqnarray}}

\newcommand\POWHEGBOX{\texttt{POWHEG BOX}}
\newcommand\POWHEG{\texttt{POWHEG}}
\newcommand\MCNLO{\texttt{MC@NLO}}
\newcommand\JETPHOX{\texttt{JETPHOX}}
\newcommand\PYTHIA{\texttt{PYTHIA}}
\newcommand\PYTHIAv{\texttt{PYTHIA 8}}
\newcommand\pwhghooks{\texttt{PowhegHooks}}
\newcommand\HERWIG{\texttt{HERWIG}}
\newcommand\FORM{\texttt{FORM}}
\newcommand\formcalc{\texttt{FormCalc}}
\newcommand\looptools{\texttt{LoopTools}}
\newcommand\madgraph{\texttt{MadGraph}}
\newcommand\madloop{\texttt{MadLoop}}
\newcommand\FORTRAN{\texttt{FORTRAN}}

\newcommand{\dd}[1]{\ensuremath{\mathrm{d}#1}}

\newcommand{\Ord}{\ensuremath{\mathcal{O}}}

\newcommand{\pT}{\ensuremath{p_\mathrm{T}}}
\newcommand{\pTmin}{\ensuremath{p_\mathrm{T}^\mathrm{min}}}
\newcommand{\kT}{\ensuremath{k_\mathrm{T}}}

\newcommand{\alphas}{\ensuremath{\alpha_\mathrm{s}}}
\newcommand{\MSbar}{\ensuremath{\overline{\text{MS}}}}
\newcommand{\LamQCD}{\ensuremath{\Lambda_\mathrm{QCD}}}
\newcommand{\Bnull}{\ensuremath{B^{(0)}}}
\newcommand{\Bone}{\ensuremath{B^{(1)}}}
\newcommand{\Btwo}{\ensuremath{B^{(2)}}}
\newcommand{\Vnull}{\ensuremath{V^{(0)}}}

\newcommand{\alr}{\ensuremath{{\alpha_\mathrm{r}}}}
\newcommand{\alrp}{\ensuremath{{\alpha'_\mathrm{r}}}}
\newcommand{\alp}{\ensuremath{{\alpha_\oplus}}}
\newcommand{\alm}{\ensuremath{{\alpha_\ominus}}}
\newcommand{\Bbar}{\ensuremath{\overline{B}}}
\newcommand{\Phibar}{\ensuremath{\overline{\Phi}}}
\newcommand{\Phirad}{\ensuremath{\Phi_\mathrm{rad}}}
\newcommand{\sumalr}{\ensuremath{\sum_{\alr \in \{\alr | f_b\}}}}
\newcommand{\sumalp}{\ensuremath{\sum_{\alp \in \{\alp | f_b\}}}}
\newcommand{\sumalm}{\ensuremath{\sum_{\alm \in \{\alm | f_b\}}}}
\newcommand{\fbqed}{\ensuremath{f_b^\mathrm{QED}}}
\newcommand{\fbqcd}{\ensuremath{f_b^\mathrm{QCD}}}

\newcommand{\kTsupp}{\ensuremath{k_\mathrm{T, supp}}}
\newcommand{\kTmin}{\ensuremath{k_\mathrm{T, \min}}}
\newcommand{\sumspincol}{\ensuremath{\sum_{\substack{\mathrm{spins}\\\mathrm{colours}}}}}
\newcommand{\Vfin}{\ensuremath{V_\mathrm{fin}}}
\newcommand{\Vsv}{\ensuremath{V_\mathrm{sv}}}
\newcommand{\calI}{\ensuremath{\mathcal{I}}}
\newcommand{\calQ}{\ensuremath{\mathcal{Q}}}
\newcommand{\calIQED}{\ensuremath{\mathcal{I}^\mathrm{QED}}}
\newcommand{\calQQED}{\ensuremath{\mathcal{Q}^\mathrm{QED}}}
\newcommand{\gammaQED}{\ensuremath{\gamma^\mathrm{QED}_{f_i}}}
\newcommand{\gammapQED}{\ensuremath{\gamma^{\mathrm{QED}'}_{f_i}}}
\newcommand{\VsvQED}{\ensuremath{V_\mathrm{sv}^{\fbqcd}}}
\newcommand{\aQED}{\ensuremath{a^\mathrm{QED}}}
\newcommand{\cijQED}{\ensuremath{c_{ij}^\mathrm{QED}}}
\newcommand{\BtwoQCD}{\ensuremath{B^{\fbqcd(2)}}}
\newcommand{\BoneQCDchij}{\ensuremath{B^{\fbqcd(1)}_{ij,\mathrm{ch}}}}
\newcommand{\BQCDchij}{\ensuremath{B^{\fbqcd}_{ij,\mathrm{ch}}}}
\newcommand{\Bchij}{\ensuremath{B_{ij,\mathrm{ch}}}}

\preprint{MS-TP-16-23}

\title{Prompt photon production and photon-hadron jet correlations with POWHEG}

\author[a]{Tom\'{a}\v{s} Je\v{z}o,}
\author[b]{Michael Klasen,}
\author[b]{and Florian K\"onig}

\affiliation[a]{Universit\'{a} di Milano-Bicocca and INFN, Sezione di Milano-Bicocca,
 Piazza della Scienza 3, I-20126 Milano, Italy}
\affiliation[b]{Institut f\"ur Theoretische Physik, Westf\"alische Wilhelms-Universit\"at
 M\"unster, Wilhelm-Klemm-Stra\ss{}e 9, D-48149 M\"unster, Germany}

\emailAdd{tomas.jezo@mib.infn.it}
\emailAdd{michael.klasen@uni-muenster.de}
\emailAdd{florian.koenig@uni-muenster.de}

\abstract{We present a calculation of direct photon production at next-to-leading
 order of QCD and a matching of this calculation with parton showers using \POWHEGBOX{}.
 Based on simulations with \POWHEG+\PYTHIA, we perform a detailed phenomenological analysis
 of PHENIX data on prompt photon production and photon-hadron jet correlations in $pp$ collisions
 at RHIC, considerably improving the description of these data with respect
 to previous calculations, and we suggest additional interesting analyses.}

\keywords{Perturbative QCD, NLO, parton showers, photons, hadron colliders}

\begin{document}
\maketitle
\flushbottom

%%%%%%%%%%%%%% Begin Section 1 %%%%%%%%%%%%%%%%%%%%%%%%%%%%%%%%%%%%%%%%%
\section{Introduction}
\label{sec:1}

In heavy-ion collisions at RHIC \cite{Adcox:2004mh,Adams:2005dq} or the LHC
\cite{Aamodt:2010pa,Aad:2010bu,Chatrchyan:2011sx}, a new state of strongly
interacting matter can be created, the so-called quark-gluon plasma (QGP).
Important probes of this hot medium of deconfined quarks and gluons are
thermal photons, which interact only electromagnetically and can thus leave
the medium without strong modifications of their thermal spectrum. The
exponential falloff of the photon spectrum at low transverse momenta ($p_T$)
can then be related to the temperature of the QGP \cite{Wilde:2012wc,%
Klasen:2013mga,Klasen:2014xra,Adam:2015lda}. The extraction of the true
temperature of the QGP at the time of its creation is complicated by the
fact that the medium is rapidly expanding and cooling, that photons are
radiated at all stages of the collision including the phases before
thermalisation and after recombination of the quarks and gluons into
charged and neutral hadrons, that neutral pions decay preferably into
pairs of photons, and that photons are also produced promptly in partonic
collisions, either directly or through fragmentation processes.

In Refs.\ \cite{Wilde:2012wc,Adam:2015lda}, a first observation of a
low-$p_T$ photon signal after subtraction of the meson decay background in
Pb-Pb collisions at the LHC has been reported by the ALICE collaboration,
and an inverse slope parameter was extracted from the $p_T$-spectrum
for 0-20\% central collisions.
Using next-to-leading order (NLO) QCD calculations, the relative
contributions to prompt-photon production from different initial and final
states and the theoretical uncertainties coming from independent variations
of the renormalisation and factorisation scales, the nuclear parton
densities and the fragmentation functions have been analysed. Based on
different fits to the unsubtracted and prompt-photon subtracted ALICE data
for 0-40\% central collisions, we found
effective temperatures of $T = 304 \pm 58$ MeV and $309 \pm 64$ MeV at
$p_T\in[0.8;2.2]$ GeV and $p_T\in[1.5;3.5]$ GeV as well as a power-law
($p_T^{-4}$) behaviour for $p_T > 4$ GeV as predicted by QCD hard scattering
\cite{Klasen:2013mga,Klasen:2014xra}. In lower-energy Au-Au
collisions at RHIC, a smaller effective temperature of $T = 221 \pm 27$ MeV
had previously been measured \cite{Adare:2008ab}.

Precise calculations of prompt photon production in hadronic collisions
are not only
imperative for a reliable extraction of the thermal photon spectrum, but
also for measurements of photon-hadron and photon-jet correlations, which
represent a second
important probe of the hot medium due to the $p_T$-imbalance and azimuthal
asymmetries induced by jet quenching \cite{Adare:2009vd,Arbor:2013ola}.
In both cases, additional parton emission can significantly modify the
spectra and thus the physical conclusions. So far, the theoretical
description of prompt photon production has relied on NLO calculations
\cite{Klasen:2002xb} and in particular \JETPHOX{} \cite{Aurenche:2006vj} with
at most one additional parton for direct and fragmentation processes. The
latter dominate at low $p_T$ and require a convolution with insufficiently
determined non-perturbative fragmentation functions \cite{Klasen:2014xfa},
unless one applies photon isolation criteria \cite{Frixione:1998jh} or
departs from real to slightly virtual photons \cite{Berger:1998ev,%
Berger:1999es,Brandt:2013hoa,Brandt:2014vva}.

An alternative approach consists in the combination of NLO calculations with
parton showers (PS). There, the photon fragmentation function can be modelled
by an interleaved QCD+QED parton shower, leading e.g.\ to a correct description
of the photon fragmentation function at LEP \cite{Hoeche:2009xc,Bellm:2015jjp}.
In addition, the exclusive events produced in this Monte Carlo approach allow
for a detailed comparison to experiment, in particular realistic isolation cuts
and even a combination with detector simulations. Furthermore, the parton
shower resums the leading logarithmic contributions from multiple additional
parton emissions, thus providing considerably more realistic kinematic
distributions. This applies in particular to photon-hadron and photon-jet correlations
with their unrealistic $\delta$-functions or divergences predicted at leading
order (LO) and NLO in the regions of vanishing photon-hadron transverse momentum
$p_T^{\gamma h}\to0$ and back-to-back azimuthal angle $\Delta\phi\to\pi$. As we
will see explicitly, the collinear region $\Delta\phi\to0$ is of course closely
related to photon fragmentation processes.

The combination of NLO QCD corrections with PS requires a careful treatment of
soft/collinear regions in order to avoid double counting. Methods like \MCNLO{}
\cite{Frixione:2002ik} and \POWHEG{} \cite{Frixione:2007vw} are now well
established for QCD processes. The treatment of photons is more intricate,
as it also requires
a QED parton shower. It has previously been achieved for di-photon production
as a Higgs boson background \cite{D'Errico:2011sd} and as an electroweak 
correction to single-$W$ production \cite{Barze:2012tt,Barze:2014zba}. In this
paper, we report on a re-calculation and validation of direct photon production
at NLO in Sec.\ \ref{sec:2}, a matching of this calculation with PS using
\POWHEGBOX{} \cite{Alioli:2010xd} in Sec.\ \ref{sec:3}, and in Sec.\
\ref{sec:4} on a successful phenomenological reanalysis of PHENIX data on
photon and photon-hadron production in $pp$ collisions at $\sqrt{s}=200$ GeV,
which form the baseline for the corresponding analyses in heavy-ion
collisions. Our conclusions are presented in Sec.\ \ref{sec:5}.

%\cite{Saimpert:2015oka}.

%Heavy quark production with pp collisions in ALICE \cite{Klasen:2014dba}\\

%Other signal of QGP formation is quarkonium suppression. Associated production of $J/\Psi$ and photons: \cite{Klasen:2004az}\\

%%%%%%%%%%%%%% Begin Section 2 %%%%%%%%%%%%%%%%%%%%%%%%%%%%%%%%%%%%%%%%%
\section{Direct photon production at NLO}
\label{sec:2}

Direct photon production proceeds at LO through the partonic processes
$q\bar{q}\to\gamma g$ and $qg\to\gamma q$. We computed the spin- and
colour-averaged cross sections for these processes analytically using
\formcalc{} 8.4 \cite{Hahn:1998yk} and checked the results against
\madgraph{} 5 \cite{Alwall:2014hca} and the literature
\cite{Berger:1983yi,Owens:1986mp,Aurenche:1987fs,Gordon:1993qc}.
The same procedure was applied to the real emission processes with
an additional parton in the final state.
The virtual one-loop corrections were computed with \FORM{}
\cite{Vermaseren:2000nd} in $D=4-2\varepsilon$ dimensions and reduced
from tensor to scalar integrals using the Passarino-Veltman procedure
\cite{Passarino:1978jh}. After renormalisation of the ultraviolet
divergences in the $\MSbar$ scheme \cite{Bardeen:1978yd}, infrared
divergences remained which could be shown to cancel against those from
the integrated Catani-Seymour dipoles \cite{Catani:1996vz} as computed
e.g.\ with \texttt{AutoDipole} 1.2.3 \cite{Hasegawa:2009tx}. For the
finite remainders of the one-loop contributions, agreement with those
produced by \madloop{} \cite{Alwall:2014hca} was then obtained.

As is well known \cite{Klasen:2002xb}, a consistent calculation of
prompt photon production up to NLO requires also the inclusion of
fragmentation processes at least in LO in order to cancel the
divergences from collinear quarks and photons. For photons with finite
transverse momenta, they appear only in the final state and are canceled
by the corresponding dipole terms arising from collinear factorisation.
The LO direct and purely partonic fragmentation
processes scale formally with $\Ord(\alpha\alphas)$ and $\Ord(\alphas^2)$,
respectively. However, the latter must still be convoluted with
fragmentation functions (FFs), which scale as
\begin{equation}
  D_{\gamma/i}(z,\mu_\gamma) \sim \alpha \ln{\frac{\mu_\gamma}{\LamQCD}} \sim \frac{\alpha}{\alpha_s}\,,
  \label{eq:FFevol}
\end{equation}
so that both contributions eventually have the same scaling behaviour.

For the numerical evaluation of our NLO direct and LO fragmentation results,
we computed the scalar loop integrals using \looptools{} 2.13
\cite{Hahn:1998yk}. The partonic cross sections were then convoluted with
parton density functions (PDFs) and FFs and compared to \texttt{JETPHOX}
\cite{Aurenche:2006vj}. As expected, good numerical agreement was found.
Important advantages of NLO over LO calculations are a more reliable
(typically larger) normalisation of the total cross section, its
stabilisation with respect to variations of the unphysical renormalisation
and factorisation scales, and improved descriptions of kinematic
distributions. Disadvantages with respect to Monte Carlo generators are
the restriction to at most one additional parton and the absence of
hadronisation effects.

%%%%%%%%%%%%%% Begin Section 3 %%%%%%%%%%%%%%%%%%%%%%%%%%%%%%%%%%%%%%%%%
\section{Prompt photon production with POWHEG}
\label{sec:3}

The \POWHEGBOX{} \cite{Alioli:2010xd} provides a framework to smoothly
incorporate NLO corrections in general-purpose event generators such as
\PYTHIA{} \cite{Sjostrand:2014zea} or \HERWIG{} \cite{Bellm:2015jjp}, as
long as they allow for a $\pT$-ordered parton
shower or have the ability to veto radiation with a $\pT$ higher than that of
the first radiation. Usually, it suffices to provide the Born amplitudes
along with their colour- and spin-correlated counterparts, the Born phase space,
a decomposition of the amplitudes in the colour flow basis,
the finite part of the virtual corrections, and the real correction
amplitudes as \FORTRAN{} routines. These inputs are then linked against the
core \POWHEGBOX{} code. In our case, however, it was necessary to modify small
parts of the \POWHEGBOX{} code itself, primarily to accommodate a consistent
treatment of photon radiation off quarks and furthermore to introduce an
artificial enhancement of photon radiation. Since the \POWHEGBOX{} in version
2 is already able to handle photon radiation off leptons as reported in Ref.\
\cite{Barze:2012tt}, only minor modifications of the code were necessary. In
the following, we use the notation established in Ref. \cite{Frixione:2007vw}.

In \POWHEG{} \cite{Frixione:2007vw}, the real processes are subdivided into
parts corresponding to collinear and soft regions, such that for a specific
flavour structure (i.e.\ partonic subprocess) the real process $R$ can be
written as a sum
\begin{equation}
  R = \sum_\alr R^\alr
  \label{eq:Rdecomp}
\end{equation}
with an index $\alr$ denoting the different singular regions. The individual
contributions $R^\alr$ are
chosen such that, for some region of the real correction phase space where the
configuration of two particles produces a collinear or soft singularity, only
one $R^\alrp$ becomes singular, while all other $R^\alr$ with $\alr \neq \alrp$
vanish. Hence, every region $\alr$ with $n+1$ particles corresponds to an
underlying Born flavour structure with $n$ particles, denoted by $f_b$ and
obtained by replacing the two particles in the singular configuration by the
particle from which they emerged in a splitting process. This defines a
mapping $\alr \to f_b$. In addition, there exists for all $\alr$ a
decomposition of the $n+1$-particle phase space, denoted by $\Phi_{n+1}$, into
$n$-particle kinematics $\Phibar_n^\alr$ and radiation variables $\Phirad$,
giving a mapping of real kinematics to Born kinematics. The decomposition in
Eq.\ \eqref{eq:Rdecomp} and the relation between $\Phibar_n^\alr$, $\Phirad$
and $\Phi_{n+1}$ are central to the Frixione-Kunszt-Signer (FKS) subtraction
method \cite{Frixione:1995ms}, which is employed by the \POWHEGBOX{} to
regularise infrared singularities, but also to the formulation of the
\POWHEG{} Sudakov form factor and the \POWHEG{} cross section. The latter
is defined as
\begin{equation}
  \begin{aligned}
    d\sigma &= \sum_{f_b} \Bbar^{f_b}(\Phi_n) \dd\Phi_n \left\{ \Delta^{f_b}(\Phi_n, \pTmin) 
      \vphantom{\sumalr \frac{\left[ \dd\Phirad \, \theta(\kT - \pTmin) \Delta^{f_b}(\Phi_n, \kT) R(\Phi_{n+1}) \right]^{\Phibar_n^\alr = \Phi_n}_\alr}{B^{f_b}(\Phi_n)}} \right. \\
    & + \left.\sumalr \frac{\left[ \dd\Phirad \, \theta(\kT - \pTmin) \Delta^{f_b}(\Phi_n, \kT) R(\Phi_{n+1}) \right]^{\Phibar_n^\alr = \Phi_n}_\alr}{B^{f_b}(\Phi_n)} \right\}\,.
  \end{aligned} 
  \label{eq:POWHEGxsec}
\end{equation}
Here, $\Bbar^{f_b}$ is the NLO inclusive $n$-particle cross section, where all
radiative corrections from regions $\alr$ with underlying Born structure $f_b$
-- denoted by the set $\{\alr | f_b\}$ -- have been integrated out, i.e.\
\begin{equation}
  \begin{aligned}
    \Bbar^{f_b}(\Phi_n) &= B^{f_b}(\Phi_n) + \Vsv^{f_b}(\Phi_n) + \sumalr \int\left[\dd\Phirad \left\{R(\Phi_{n+1}) - C(\Phi_{n+1})\right\}\right]^{\Phibar_n^\alr = \Phi_n}_\alr \\
    & + \sumalp \int\frac{\dd z}{z} G^\alp_\oplus(\Phi_{n,\oplus}) + \sumalm \int\frac{\dd z}{z} G^\alm_\ominus(\Phi_{n,\ominus})
  \end{aligned}
  \label{eq:Bbar}
\end{equation}
with the Born amplitude $B$, the subtracted virtual corrections $\Vsv$, the
real counterterms $C$ and the collinear remnants $G_\oplus$ and $G_\ominus$. 

Eq.\ \eqref{eq:POWHEGxsec} is the cross section for events with at most one
radiation off the Born flavour structure with $\pT > \pTmin$, or in parton
shower terminology the cross section with the first radiation evolved down
to $\pTmin$. Compared to the usual parton shower prescription for the first
radiation, the \POWHEG{} modifications are given by the replacement $B \to
\Bbar$ and the substitution of the parton shower splitting kernel (usually
the Altarelli-Parisi splitting kernel) with the ratio of the real and Born
amplitudes $R^\alr(\Phi_{n+1})|^{\Phibar_n^\alr = \Phi_n}/B^{f_b}(\Phi_n)$ for each
radiation region $\alr \in \{\alr | f_b\}$. Consequently, the \POWHEG{}
Sudakov form factor is given by
\begin{equation}
  \Delta^{f_b}(\Phi_n, \pT) = \exp\left\{- \sumalr \int\frac{\left[\dd\Phirad R(\Phi_{n+1}) \theta(\kT(\Phi_{n+1}) - \pT)\right]^{\Phibar_n^\alr = \Phi_n}_\alr}{B^{f_b}(\Phi_n)}\right\}\,.
  \label{eq:POWHEGsuda}
\end{equation}

In our treatment of photon production at NLO we follow the approach of Ref.\
\cite{D'Errico:2011sd} and include all real amplitudes $R$ with one photon
in the final state. Our goal is then to simulate the fragmentation
contribution arising from the QED radiation off partons (as described for a
tree-level merging approach in Ref.\ \cite{Hoeche:2009xc}). Thus, we need to
include the Born flavour structures $\fbqcd$ for LO dijet production
along with the photon production flavour structures $\fbqed$ as underlying
Born processes. In this case the sum over the flavour structures $\fbqcd$ in
Eq.\ \eqref{eq:POWHEGxsec} includes the photon fragmentation contribution
through the combination of QED non-branching and branching probabilities,
i.e.\ the terms in the brackets.

%%%%%%%%%%%%%% Begin Section 3.1 %%%%%%%%%%%%%%%%%%%%%%%%%%%%%%%%%%%%%%%
\subsection{Born amplitudes and phase space}
\label{sec:born}

As described above, we implement both the QED and QCD Born amplitudes, expanded
in $D = 4 - 2 \varepsilon$ dimensions up to $\Ord(\varepsilon^2)$ for reasons
specified in Sec.\ \ref{sec:virt}. Aside from the rather straight-forward implementation
of the Born amplitudes themselves, an assignment of colours to the external
legs has to be given according to the large-$N_c$ limit. To facilitate this,
one identifies the squared diagrams for a specific amplitude which have a
planar colour flow and assigns a colour to each of their lines. If then several
diagrams have conflicting colour assignments, one is chosen according to the
relative weight of its planar diagram with respect to the sum of all planar
diagrams. However, in the case of QED Born amplitudes this is not necessary,
since there is only one $q\bar{q}g$-vertex. It is therefore trivial to assign a
colour to each quark. For the more complicated case of the QCD Born amplitudes,
we use the colour assignment of the \POWHEGBOX{} implementation of jet pair
production \cite{Alioli:2010xa}.

Since we treat all particles as massless, the Born phase space is the same for
our process and dijet production, enabling us to also use the phase space
routines from that code. As the Born amplitudes diverge for vanishing squared
partonic momentum transfer $\hat{t}$, a phase space cut on the final state
transverse momentum $\kT$ with respect to the beam axis is mandatory.
To regularise the divergence, we make
use of the Born suppression factor \cite{Alioli:2010xa}
\begin{equation}
  \mathcal{S}(\kT) = \left(\frac{\kT^2}{\kT^2 + \kTsupp^2}\right)^i\,.
  \label{eq:bornsuppfact}
\end{equation}
This feature is activated in the parameter file by setting the option
\texttt{bornsuppfact} to a value for $\kTsupp$ and has the effect of replacing
the functions $\Bbar^{f_b}(\Phi_n)$ in Eq.\ \eqref{eq:POWHEGxsec} with
$\mathcal{S}(\kT)\Bbar^{f_b}(\Phi_n)$. One thus avoids the divergence of the
Born amplitudes. This change of the cross section is corrected for by
weighting the generated events with the inverse of Eq-\ \eqref{eq:bornsuppfact}.
By default, we use $\kTsupp=100$ GeV and $i=3$ in Eq.\ \eqref{eq:bornsuppfact},
but we have checked (with $\kTsupp=10$ GeV and $i=4$) that within statistics
and at sufficiently large $p_T$ our results are independent of these choices.
Alternatively, a simple cut on $\kT$ can be activated with the parameter \texttt{bornktmin}.
After showering, this cut would, however, lead to a loss of events in the
low-$p_T$ region, which is precisely the region in heavy-ion collisions,
where one wants to extract the thermal photon spectrum.
We have instead tested the continuous approximation of the Heaviside function
\beq
 S(k_T)=\frac{1}{\pi} \left[\arctan[(\kT-\kTmin)\cdot10^4]+\frac{\pi}{2}\right]
 \simeq\Theta(\kT-\kTmin),
\eeq
which does, however, not improve significantly on the statistics.

Independently of the method used to regulate the LO divergence, it must be noted
that the impact of a choice of $\kTsupp$ (or a $\kT$ cut) on the photon
spectrum is not obvious, since $\kT$ is not generally the photon transverse
momentum, but the momentum of some particle prior to any generation of
radiation.
In particular, in the original \POWHEG{} scheme, only the splitting $q\to q\gamma$,
but not $q\to\gamma q$, was generated, which could yield high-$p_T$ photon events
with low statistics, but large weights after showering. Adding the line
\texttt{doublefsr 1} to the file \texttt{powheg.input} allows to generate also
the splitting $q\to\gamma q$, which avoids this statistical problem
\cite{Nason:2013uba}.

%%%%%%%%%%%%%% Begin Section 3.2 %%%%%%%%%%%%%%%%%%%%%%%%%%%%%%%%%%%%%%%
\subsection{Colour-correlated Born amplitudes}
\label{sec:colcor}

In the soft limit, a real correction is given by the Born amplitude times
an eikonal factor, where the latter depends on the colour correlations
between coloured legs of the Born amplitude. To enable the \POWHEGBOX{}
to compute these limits, the colour-correlated Born amplitudes, defined by
\begin{equation}
  B_{ij} = - N \sumspincol \mathcal{M}_{\{c_k\}} \left(\mathcal{M}_{\{c_k\}}^\dagger\right)_{\substack{c_i \to c_i'\\c_j \to c_j'}} T^a_{c_i,c_i'} T^a_{c_j,c_j'}\,,
  \label{eq:Borncolcor}
\end{equation}
are given as an input. In Eq.\ \eqref{eq:Borncolcor}, the Born matrix
elements are denoted by $\mathcal{M}_{\{c_k\}}$ with the colours of the
external particles specified by the index set $\{c_k\}$ and the averaging
factors subsumed in $N$. The colour indices for the legs $i$ and $j$ are
contracted with colour charge operators $T^a$.

The colour correlations for processes with less than four coloured partons
-- as in our QED Born processes -- are readily reduced to sums of Casimir
operators by making use of the colour conservation relations
\begin{equation}
  \begin{aligned}
    \sum_i T^a_{bc_i} \mathcal{M}_{\{c_k\}}^\dagger = 0\,,\\
    \sum_i \mathcal{M}_{\{c_k\}} T^a_{c_ib} = 0\,,
  \end{aligned}
  \label{eq:colcon}
\end{equation}
with the sum running over all coloured legs $i$ and an implicit contraction
of the colour index $c_i$. Physically, these equations simply state that an
infinitesimal global transformation of all initial and final state colours
simultaneously has no effect. From this it follows that
\begin{equation}
  \sum_{i,i \neq j} B_{ij} = C_{f_j} B
  \label{eq:Borncas}
\end{equation}
with the Casimir constants $C_{f_j}$.

Eq.\ \eqref{eq:Borncas} gives three equations for three
coloured legs, which can be solved for all the
colour correlations, since $B_{ij}$ is symmetric under the exchange of $i$
and $j$.
Thus for $q \bar{q} \to \gamma g$ with momenta $p_1$, $p_2$, $p_3$ and $p_4$,
respectively, we have
\begin{align}
  B_{12} &= \frac{2 C_F - C_A}{2} B\,, 
  \label{eq:B12} \\
  B_{14} &= B_{24} = \frac{C_A}{2} B\,.
  \label{eq:B14}
\end{align}
The results for all other partonic QED subprocesses are related to this result
via crossing symmetry. The colour correlations for the QCD Born flavour
structures are not needed, since we do not implement $\Ord(\alphas^3)$ corrections.

%%%%%%%%%%%%%% Begin Section 3.3 %%%%%%%%%%%%%%%%%%%%%%%%%%%%%%%%%%%%%%%
\subsection{Spin-correlated Born amplitudes}
\label{sec:spincor}

Similarly to the colour correlations for the construction of soft limits,
spin-correlated amplitudes are required to construct the collinear limits for
gluons splitting into two partons. The spin-correlated Born amplitude is
defined via
\begin{equation}
  B_j^{\mu\nu} = N \sumspincol \mathcal{M}_{\{s_k\}} \left(\mathcal{M}^\dagger_{\{s_k\}}\right)_{s_j \to s_j'} (\varepsilon_{s_j}^\mu)^\ast \varepsilon_{s_j'}^\nu\,,
  \label{eq:Bornspincor}
\end{equation}
where now the index $s_j$ represents the spin of the gluon on leg $j$ and
$\varepsilon_{s_j}^\mu$ is a polarisation vector for leg $j$. This prescription
amounts to replacing the polarisation vectors of leg $j$ in the matrix
element $\mathcal{M}$ and its Hermitian conjugate $\mathcal{M}^\dagger$ with
the physical polarisation sums, e.g.
\begin{equation}
  \sum_{s_j} \varepsilon_{s_j}^\rho (\varepsilon_{s_j}^\mu)^\ast = -g^{\mu\rho} + \frac{p_j^\mu \eta^\rho + p_j^\rho \eta^\mu}{p_j \cdot \eta}\,,
  \label{eq:polsum}
\end{equation}
with $\eta$ some light-like vector spanning Minkowski space together with
$p_j$ and the two physical polarisations $\varepsilon_{s_j}$. As is usual, the
polarisation vectors are chosen to be space-like, orthogonal, and normalised
to unity,
\begin{equation}
  g_{\mu\nu} (\varepsilon_{s_j}^\mu)^\ast \varepsilon_{s_j'}^\nu = -\delta_{s_js_j'}\,,
  \label{eq:polnorm}
\end{equation}
leading to 
\begin{equation}
  g_{\mu\nu} B _j^{\mu\nu} = -B\,.
  \label{eq:spincorconsistence}
\end{equation}

Taking again the partonic process $q \bar{q} \to \gamma g$ and performing
the described substitutions of polarisation vectors for the gluon on leg
$4$ in \FORM{}, we find
\begin{equation}
  B_4^{\mu\nu} = \frac{1}{2} \left(-g^{\mu\nu} + \frac{p_4^\mu \eta^\nu + p_4^\nu \eta^\mu}{p_4 \cdot \eta}\right) B\,.
  \label{eq:B4munu}
\end{equation}
In practice, we choose $\eta^\mu = g^{\mu\mu} p_4^\mu$ (no sum over $\mu$).
Again, the other QED processes are obtained using crossing symmetry.
Here, again, the correlations for QCD Born amplitudes are not needed, since
only QED radiation is allowed off them. We also do not need the spin
correlations for photons, since they are not allowed to split, i.e.\ we do
not include real corrections with an internal photon line.

%%%%%%%%%%%%%% Begin Section 3.4 %%%%%%%%%%%%%%%%%%%%%%%%%%%%%%%%%%%%%%%
\subsection{Virtual corrections}
\label{sec:virt}

The virtual corrections are provided to the \POWHEGBOX{} in the form of the
finite part $\Vfin$ of the $\MSbar$-renormalised virtual amplitude $V$. The
relation between $V$ and $\Vfin$ is given in the conventional dimensional
regularisation (CDR) scheme via
\begin{equation}
  V = \frac{(4\pi)^\varepsilon}{\Gamma(1-\varepsilon)} \frac{\alphas}{2\pi} \left[ \frac{1}{\varepsilon^2} a B + \frac{1}{\varepsilon} \sum_{i,j} c_{ij} B_{ij} + \Vfin \right]\,.
  \label{eq:VCDR}
\end{equation}
While the coefficients $a$ and $c_{ij}$ are independent of $\varepsilon$,
the amplitudes $B$, $B_{ij}$ are given in $D=4-2\varepsilon$ dimensions and
thus depend on $\varepsilon$. Since our results computed in Sec.\ \ref{sec:2}
have the form
\begin{equation}
  V = (4\pi)^\varepsilon \frac{\Gamma^2(1-\varepsilon)\Gamma(1+\varepsilon)}{\Gamma(1-2\varepsilon)} \left[ \frac{1}{\varepsilon^2} V^{(-2)} + \frac{1}{\varepsilon} V^{(-1)} + \Vnull \right]\,,
  \label{eq:myV}
\end{equation}
with all $\varepsilon$-dependence explicit, it was necessary to compute from
our finite result $\Vnull$ the CDR-finite result $\Vfin$. By expanding the
Born amplitudes in $\varepsilon$,
\begin{equation}
  B = \Bnull + \varepsilon \Bone + \varepsilon^2 \Btwo + \Ord(\varepsilon^3)\,,
  \label{eq:Bornepsexp}
\end{equation}
and comparing Eqs.\ \eqref{eq:VCDR} and \eqref{eq:myV}, it is easy to see that
the relation
\begin{equation}
  \Vfin = \frac{2\pi}{\alphas} \Vnull - c_{ij} \Bone_{ij} - a \Btwo
  \label{eq:myVtoCDR}
\end{equation}
holds, which we implement in the code. The coefficients $a$ and $c_{ij}$ can be
extracted from Ref.\ \cite{Frixione:1995ms}, giving
\begin{align}
  a &= - \sum_i C_{f_i}\,,
  \label{eq:coeffa} \\
  c_{ij} &= (1-\delta_{ij}) \left[ -\frac{\gamma_{f_i}}{C_{f_i}} + \ln\left(\frac{2 p_i \cdot p_j}{\mu_R^2}\right) \right]\,,
  \label{eq:coeffcij}
\end{align}
where $i,j$ run over all coloured legs and $\mu_R$ is the renormalisation scale.
The constants $\gamma_{f_i}$ are given by
\begin{align}
  \gamma_{q,\bar{q}} &= \frac{3}{2} C_F\,,
  \label{eq:gammaq}\\
  \gamma_g &= \frac{11}{6}C_A - \frac{2}{3} T_F N_f\,,
  \label{eq:gammag}
\end{align}
where $T_F = 1/2$ as usual.

We implement the virtual $\Ord(\alphas)$ corrections to the photon production
processes using Eq.\ \eqref{eq:myVtoCDR}, but do not include the virtual
$\Ord(\alphas)$ and $\Ord(\alpha)$ corrections to the dijet processes,
as they lead to higher-order corrections for prompt photon production. We
comment on the cancellation of divergences and the inclusion of finite remnants
of the subtraction method in the next section.

%%%%%%%%%%%%%% Begin Section 3.5 %%%%%%%%%%%%%%%%%%%%%%%%%%%%%%%%%%%%%%%
\subsection{Real corrections and the cancellation of divergences}
\label{sec:realdiv}

In our implementation of the real corrections with a final state photon,
we have to make sure that the QED collinear and soft divergences are correctly
identified. In order for the \POWHEGBOX{} to find the QED singularities in
addition to the QCD singularities, it has to treat the photon as just another
massless parton, which is achieved by setting the variable
\texttt{flst\_lightpart} to $3$. Then all singular regions $\alr$ of the real
amplitudes are automatically identified and subtracted. As a check of consistency
of the real corrections and the colour- and spin-correlated Born amplitudes from
Secs.\ \ref{sec:colcor} and \ref{sec:spincor}, the \POWHEGBOX{} tests numerically
if, for each singular limit, the ratio of $R^\alr$ and the corresponding limiting
expression tends to one. Our implementation passes these tests.

The subtraction of singularities is handled internally by the \POWHEGBOX{} with
the FKS subtraction method. The soft and collinear counterterms (denoted by $C$
in Eq.\ \eqref{eq:Bbar}) are automatically assembled from the soft and collinear
limiting expressions and subtracted from the real corrections. On the other hand,
the counterterms, integrated in $D=4-2\varepsilon$ dimensions over the momentum of
the emitted parton, are added to the virtual corrections $V$ in Eq.\
\eqref{eq:VCDR}, leading to a cancellation of poles apart from collinear
initial-state singularities, which are automatically absorbed into parton
distribution functions in the $\MSbar$ factorisation scheme. The leftovers from
the absorption of the divergences of the initial-state collinear counterterms into
the PDFs are the collinear remnants $G_\oplus$ and $G_\ominus$ in Eq.\ \eqref{eq:Bbar}.
Since these are automatically computed by the \POWHEGBOX{} for all Born flavour
structures, we implemented {\em if}-clauses, which disable the (in our case
inconsistent) computation of these terms for the $\fbqcd$ amplitudes.

However, even apart from the collinear initial-state singularities, the addition
of integrated counterterms and virtual corrections does not equal $\Vfin$. Rather
there are some finite terms, that are computed automatically by the \POWHEGBOX{}
giving the {\em soft-virtual} contribution
\begin{equation}
  \Vsv = \frac{\alphas}{2\pi} \left(\calQ B + \sum_{ij} \calI_{ij} B_{ij} + \Vfin\right)
  \label{eq:softvirt}
\end{equation}
that enters Eq.\ \eqref{eq:Bbar}. The definitions of $\calI$ and $\calQ$ are
provided in Ref.\ \cite{Frixione:2007vw}. Here, we again made sure that these
terms are only computed for the $\fbqed$ amplitudes, i.e.\ those amplitudes for
which we actually implemented $\Vfin$.

Since we do not implement the virtual QED corrections to the $\fbqcd$ flavour
structures, as would be required in a fully consistent treatment, there is some
ambiguity in the choice of the finite remnants of the soft and collinear QED
singularities. In a fixed-order calculation, this freedom of choice amounts to the
choice of a factorisation scheme for the photon fragmentation function. In analogy
to the $\MSbar$ factorisation scheme, we cancel only the poles of the QED
singularities, which (by inspection of Eq.\ \eqref{eq:myV}) is equivalent to setting
$\Vnull = 0$ in the adaptation of Eq.\ \eqref{eq:myVtoCDR} for QED corrections. The
implementation of the QED version of Eq.\ \eqref{eq:softvirt} then accounts for the
finite terms in the QED counterterms. 

The contribution of Eq.\ \eqref{eq:softvirt} for QED corrections can be activated
by setting the flag \texttt{flg\_with\_em = .TRUE.} in the code, whose function we
have extended to massless quarks. With the
$\calI$-terms for QED already included in the \POWHEGBOX{} version 2, we just added
the $\calQ$-term for photon radiation off massless quarks with
\begin{equation}
  \begin{aligned}
    \calQ^\mathrm{QED} &= \sum_i \left[\gammapQED - \ln\left(\frac{s}{\mu_R^2}\right) \left(\gammaQED - 2 Q_{f_i}^2 \ln\left(\frac{2 E_i}{\xi_c \sqrt{s}}\right)\right)\right. \\
    &\left.+ 2 Q_{f_i}^2 \left(\ln^2\left(\frac{2 E_i}{\sqrt{s}}\right) - \ln^2(\xi_c)\right) - 2 \gammaQED \ln\left(\frac{2 E_i}{\sqrt{s}}\right)\right]
    \label{eq:Qqed}
  \end{aligned}
\end{equation}
(cf.\ Eq.\ (2.100) in Ref.\ \cite{Frixione:2007vw} with $\delta_o = 2$),
where the sum is over all charged legs, $Q_{f_i}$ is the charge of particle $f_i$,
and
\begin{align}
  \gammaQED &= \frac{3}{2} Q_{f_i}^2\,, \\
  \gammapQED &= \left(\frac{13}{2} - \frac{2 \pi^2}{3}\right) Q_{f_i}^2
\end{align}
(cf.\ Eq.\ \eqref{eq:gammaq}).

In summary, our soft-virtual term for the QED singularities is
\begin{equation}
 \VsvQED = \frac{\alphas}{2\pi} \left(\calQQED B^{\fbqcd} + \sum_{ij} \calIQED_{ij} \BQCDchij 
 - \aQED \BtwoQCD - \cijQED \BoneQCDchij \right)\,,
  \label{eq:softvirtQED}
\end{equation}
with
\begin{align}
  \aQED &=  - \sum_i Q_{f_i}^2\,,\\
  \cijQED &= (1-\delta_{ij}) \left[ -\frac{\gammaQED}{Q^2_{f_i}} + \ln\left(\frac{2 p_i \cdot p_j}{\mu_R^2}\right) \right]
\end{align}
(cf.\ Eqs.\ \eqref{eq:coeffa} and \eqref{eq:coeffcij}).
The charge-correlations (which are already present in the \POWHEGBOX{} version 2)
are defined in analogy to Eq.\ \eqref{eq:Borncolcor} as
\begin{equation}
  \Bchij = - B Q_{f_i} Q_{f_j} (-1)^{\sigma_i + \sigma_j}\,,
\end{equation}
where $\sigma_i = 0$, if $f_i$ is a (initial-) final-state (anti-) particle,
and $\sigma_i = 1$, if it is a (final-) initial-state (anti-) particle.

%%%%%%%%%%%%%% Begin Section 3.6 %%%%%%%%%%%%%%%%%%%%%%%%%%%%%%%%%%%%%%%
\subsection{Enhanced QED radiation}

As it stands, the implementation of single-photon production in the
\POWHEGBOX{} framework described in the preceding sections leads to a very
low photon production rate. For example, a $pp$-run at the PHENIX energy
$\sqrt{s} = 200$ GeV contains a photon in only about $2\%$ of the events,
while the remaining events are made up of QCD Born configurations.
Reasons for this behaviour are a relative suppression of photonic vs.\
purely partonic processes from the ratio of electromagnetic and strong
coupling constants $(\alpha/\alpha_s)$, QCD colour factors larger than
unity vs.\ squared fractional quark charges smaller than unity, and a
larger multiplicity of contributing processes in QCD.
To boost the contribution of photons, we implement a procedure described in
Ref.\ \cite{Hoeche:2009xc}: We multiply the integrand in the exponent of
the \POWHEG{} Sudakov form factor in Eq.\ \eqref{eq:POWHEGsuda}, denoted
by $f(\Phirad)$ in the following, for QED radiation with a constant
$c > 1$, thus decreasing the no-branching probability, and compensate for it
by reweighting the event.

Usually, the transverse momentum of a radiation is generated by first
solving the equation
\begin{equation}
  \Delta^{(U)}(\pT) = r
  \label{eq:genpt}
\end{equation}
for a uniformly distributed random number $r \in [0,1]$, where $\Delta^{(U)}(\pT)$
is a lower bound for Eq.\ \eqref{eq:POWHEGsuda}, obtained by replacing $f(\Phirad)$
by an upper bounding function $U(\Phirad) > f(\Phirad)$. Afterwards, the generated
$\pT$ is accepted with a probability given by the ratio $f/U$ or vetoed with a
probability $1 - f/U$. In the latter case Eq.\ \eqref{eq:genpt} is solved again,
but this time restricting $\pT$ to values below the vetoed value. The whole
procedure is then reiterated, until a $\pT$ is accepted or when
a low $\pT$-cutoff of the order of $\LamQCD$ is reached. Replacing Eq.\
\eqref{eq:genpt} with
\begin{equation}
  \ln\left(\Delta^{(U)}(\pT)\right) = \frac{\ln(r)}{c}
  \label{eq:genptmod}
\end{equation}
has in the \POWHEGBOX{} the same effect as multiplying both $f$ and $U$ by $c$. As described in
App.\ B of Ref.\ \cite{Hoeche:2009xc}, we then compensate for this by weighting the
event with 
\begin{equation}
  w = \frac{1}{c} \prod_i \frac{1 - \frac{f_i}{c U_i}}{1 - \frac{f_i}{U_i}},
  \label{eq:reweight}
\end{equation}
where the product runs over all QED radiation vetoes and $f_i$, $U_i$ are the
values of $f$ and $U$ at the respective vetoed $\pT$.
A similar procedure, described in Ref.\ \cite{Lonnblad:2012hz}, has also been
tested and produces consistent results.

%%%%%%%%%%%%%% Begin Section 3.7 %%%%%%%%%%%%%%%%%%%%%%%%%%%%%%%%%%%%%%%
\subsection{Parton shower with \PYTHIAv{}}

The events generated by the \POWHEGBOX{} have to be passed to a parton shower
generator to produce complete events. Every parton shower generator that is
$\pT$-ordered or has facilities to veto radiation with a $\pT$ higher than the
scale of the hardest event is viable. \PYTHIA{} is employing a $\pT$-ordered
parton shower, and we use \PYTHIAv{} \cite{Sjostrand:2014zea} in this work.
However, since the definitions of the transverse momentum of a radiation
differ for the \POWHEGBOX{} and \PYTHIA{}, it is suggested to use the class
\pwhghooks{} to account for the differences. The preferred mode of usage is to
have \PYTHIA{} evolve the shower starting from the kinematical limit rather
than the scale passed by the \POWHEGBOX{}, translate the $\pT$ of a generated
radiation from the \PYTHIA{} definition to the \POWHEGBOX{} definition, and
then veto radiation harder than the hard \POWHEG{} scale.

In addition, there is in our case the question of how to handle the scales
for QED and QCD radiation. The default would be to make no distinction and veto
the evolution of QED and QCD radiation above the \POWHEG{} scale independently
of the type of event. Instead, we follow the approach presented in Ref.\
\cite{D'Errico:2011sd}, which suggests employing two different hard scales, one
for the QED and one for the QCD shower. A discussion of this approach and another
approach, that includes a competition between QED and QCD radiation already at
the level of the hard process (i.e.\ an implementation of $R\sim\Ord(\alpha_s^3)$),
can be found in
Ref.\ \cite{Barze:2014zba}. In the nomenclature of that reference, we use the
NC-scheme.

To allow for the distinctive treatment of QED and QCD radiation, we modify
the \POWHEGBOX{} code to pass the scale of the underlying Born event, in addition
to the scale of the first radiation, to \PYTHIA{}. According to our
implementation, the Born scale is the hard scale for QED radiation, if the
underlying Born event includes a photon, or the hard scale for QCD radiation, if
the underlying Born event is a pure QCD event. The scale of the emission that is
usually passed by the \POWHEGBOX{} corresponds in those two cases to the hard
scale for QCD and QED radiation, respectively. By a modification of the
\pwhghooks{} class, we ensure that the QED and QCD showers are vetoed accordingly.

%%%%%%%%%%%%%% Begin Section 4 %%%%%%%%%%%%%%%%%%%%%%%%%%%%%%%%%%%%%%%%%
\section{Comparison with PHENIX data}
\label{sec:4}

In this section, we present the numerical results of our implementation of
the NLO corrections to prompt photon production into the \POWHEGBOX{}
matched to the \PYTHIAv{} parton shower.\footnote{The new version of \POWHEG{}
is available from the authors upon request.} We take the opportunity to reanalyse
data taken by the PHENIX collaboration in $pp$ collisions at $\sqrt{s}=200$
GeV on nearly-real virtual and real inclusive photons at low ($p_T\in[1;6]$ GeV
and $p_T\in[5;16]$ GeV, respectively) \cite{Adare:2012vn} and higher transverse
momenta ($p_T\in[5;25]$ GeV) \cite{Adare:2012yt}. These ``vacuum'' data are not
only important as a baseline for heavy-ion collisions, but the low-$p_T$ region
is also interesting for studies of perturbative QCD itself, in particular
of the fragmentation contribution, photon-jet and photon-hadron correlations,
and soft radiation effects.
Besides comparing our new results to data, we also validate them against
a pure NLO calculation with \JETPHOX{}, pointing out important similarities
and differences, i.e.\ the common NLO normalisation, but the presence of
only one additional parton and of fragmentation contributions parameterised
by non-perturbative fragmentation functions in the latter.
In our comparison with the stand-alone \PYTHIAv{} Monte Carlo generator,
we emphasise the common multiple parton emission, which leads to a better
description of kinematic distributions, and the different (NLO vs.\ LO)
normalisation.

% PDFs, FFs, Scales

In all our theoretical calculations, we employ CT14NLO parton densities
in the proton \cite{Dulat:2015mca}. Proton PDFs including photons and
photon radiation through LO QED evolution are also available
\cite{Schmidt:2015zda}. However, we did not use these for several reasons:
they pertain to only initial and initial-state soft and collinear photons,
would require the implementation of the full
(i.e.\ also virtual) QED corrections, affect the production of
prompt photons with finite $p_T$ only beyond LO in QED, and are
thus here of little numerical importance. In our comparison
with \JETPHOX{}, we employ the BFG II photon fragmentation functions
\cite{Bourhis:1997yu}, which we have previously shown to be favoured by
the PHENIX low-$p_T$ data \cite{Klasen:2014xfa}. By default, the
renormalisation and factorisation scales are set to the $p_T$ of the
underlying Born process, which can be a parton for LO QCD processes.
%Scale uncertainties are estimated by varying scales by relative factors of two, but not four.
%
% Other parameters
%
As mentioned in the previous section, we have used the Born suppression factor
in Eq.\ \eqref{eq:bornsuppfact} with $\kTsupp=100$ GeV and $i=3$ by setting
\texttt{bornsuppfact 100}. No
generation cut on the Born $k_T$ is applied.
QED radiation is enhanced from $\sim 2~\%$ to $\sim 25~\%$ through Eq.\
\eqref{eq:genptmod} and Eq.\ \eqref{eq:reweight} with $c=50$, and independence
of the precise value of $c$ has been checked by varying it to $c=100$.

% Experiment

The PHENIX experiment has detected real photons with two electromagnetic
calorimeter (EMCal) arms covering the pseudorapidity range $|\eta^\gamma|<0.35$.
Conversion photons were identified with additional electron hits in the
ring imaging Cerenkov detector. During the 2006 RHIC runs, integrated
luminosities of ${\cal L} = 4.0$ and $8.0$ pb$^{-1}$ were collected in
the low \cite{Adare:2012vn} and higher $p_T$ ranges \cite{Adare:2012yt}.
After subtracting the decay background, the prompt photon with the hardest
$p_T$ was selected in each event.
In the higher-$p_T$ analysis, also the effects of an isolation cut on
the photons was analysed. There, the hadronic energy fraction in a cone
of radius $R=\sqrt{(\Delta\eta)^2+(\Delta\phi)^2}\leq0.5$ was restricted
to be less than 10\% of the photon energy.
%\item Ratio isolated/all photons: Fig.\ 13 and Tab.\ 3
%
In a previous publication, the PHENIX collaboration have published data
on photon-hadron jet correlations \cite{Adare:2009vd}. Using integrated luminosities
of 3.0 and 10.7 pb$^{-1}$ collected during 2005 and 2006 RHIC runs, they
identified charged-hadron jets with a tracking system composed of a drift
chamber and pixel pad chambers. Like the PHENIX collaboration, we take into
account in our calculations
all charged hadrons associated with a photon and within the PHENIX acceptance.
We have checked that modeling the charged-hadron jet in a modern way with
the anti-$k_T$ cluster algorithm and a distance parameter $R=0.4$ and rejecting
events with jets that do not contain a charged hadron produces similar results.

\subsection{Transverse-momentum spectrum of prompt photons}

\begin{figure}[!h]
 \includegraphics[width=\linewidth]{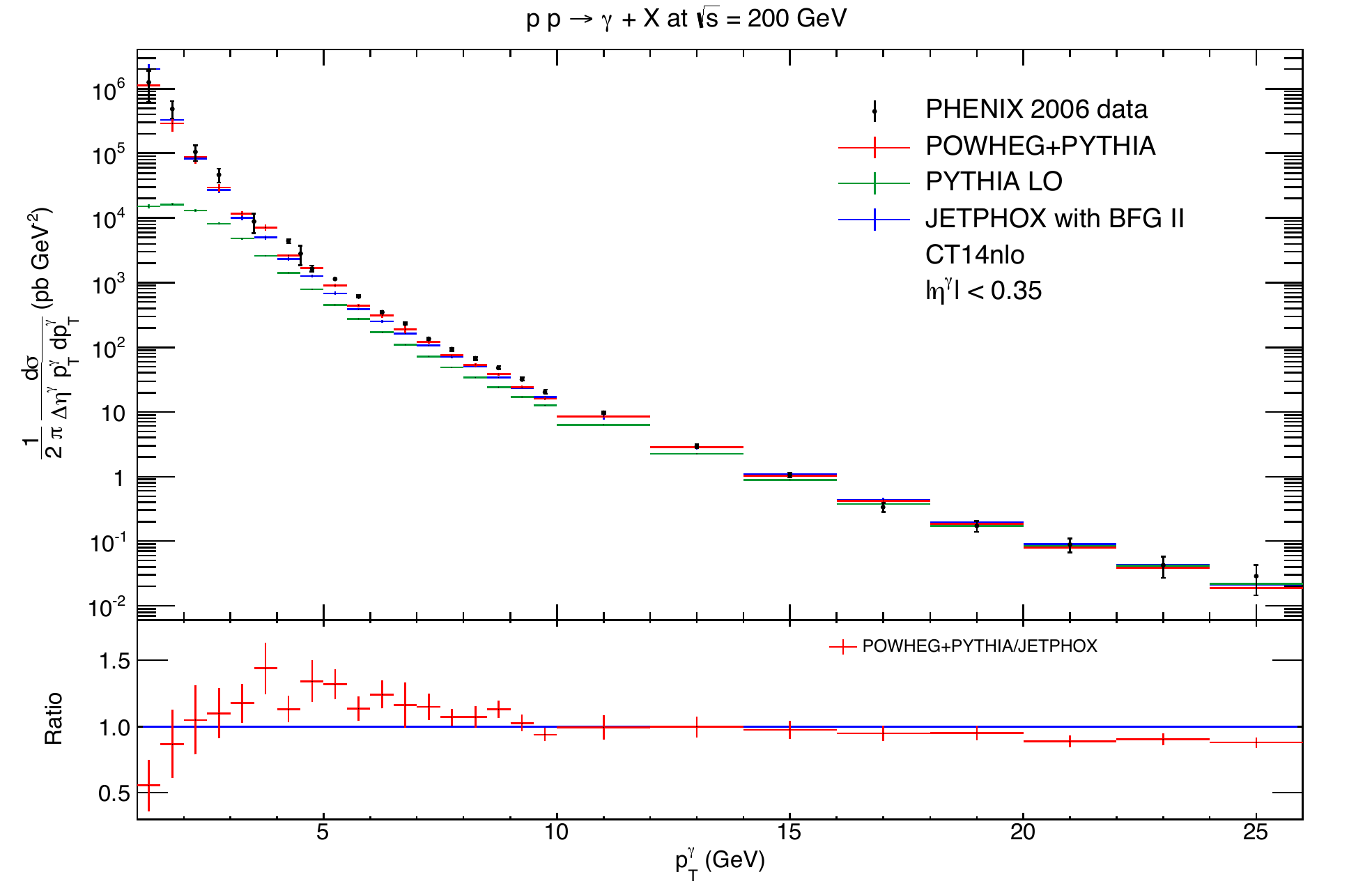}
 \caption{\label{fig:1}Transverse-momentum spectrum of prompt photons in $pp$
 collisions with $\sqrt{s}=200$ GeV at RHIC in LO+PS (green), NLO (blue) and
 NLO+PS (red) and compared to PHENIX data (black) \cite{Adare:2012vn,Adare:2012yt}.
 The ratio of NLO+PS over pure NLO is shown in the lower panel.}
\end{figure}

We begin the discussion of our numerical results with the transverse-momentum
spectrum of prompt photons in $pp$ collisions with $\sqrt{s}=200$ GeV at RHIC,
shown in Fig.\ \ref{fig:1}. 
%% all plots based on 450 datasets (i.e. different seeds) with 400k events
%% datasets not used in Fig. 1 and 1b, due to statistical instabilities:
%% 0002 0009 0010 0011 0012 0013 0014 0015 0016 0017 0075 0114 0136 0147
%% 0160 0167 0193 0194 0195 0196 0197 0207 0208 0209 0210 0211 0220 0221
%% 0222 0223 0224 0225 0226 0284 0295 0299 0300 0305 0314 0349 0402 0411
%% 0446 0447
%
As one can see, the LO+PS prediction with \PYTHIA{} alone (green) starts to
describe the data only at $p_T^\gamma>14$ GeV, but falls short of the experimental
cross section below this value and by up to two orders of magnitude in the
lowest bin. The NLO prediction with \JETPHOX{}, the fragmentation function
BFG II and central scale choices ($\mu_R=\mu_p=\mu_\gamma=p_T$, blue) describes
the measured inclusive photon spectrum reasonably well as expected.\footnote{A better
description of the data is obtained with the scale choices $\mu_R=\mu_p=p_T/2,\mu_\gamma=2p_T$
\cite{Klasen:2014xfa}.} This is even more true for the NLO+PS prediction with
\POWHEG{}+\PYTHIA{} (red), which coincides with the NLO prediction within
statistical errors over a wide range of $p_T>10$ GeV, thus validating the
calculation in a region that should be insensitive to multiple soft/collinear
parton emissions. At lower $p_T$, where these emissions become relevant, the
NLO+PS prediction exhibits a characteristic increase (lower
panel of Fig.\ \ref{fig:1}) and follows the data very well, while the pure NLO
prediction over-/undershoots the lowest and second-lowest data points.

\begin{figure}[!h]
 \includegraphics[width=\linewidth]{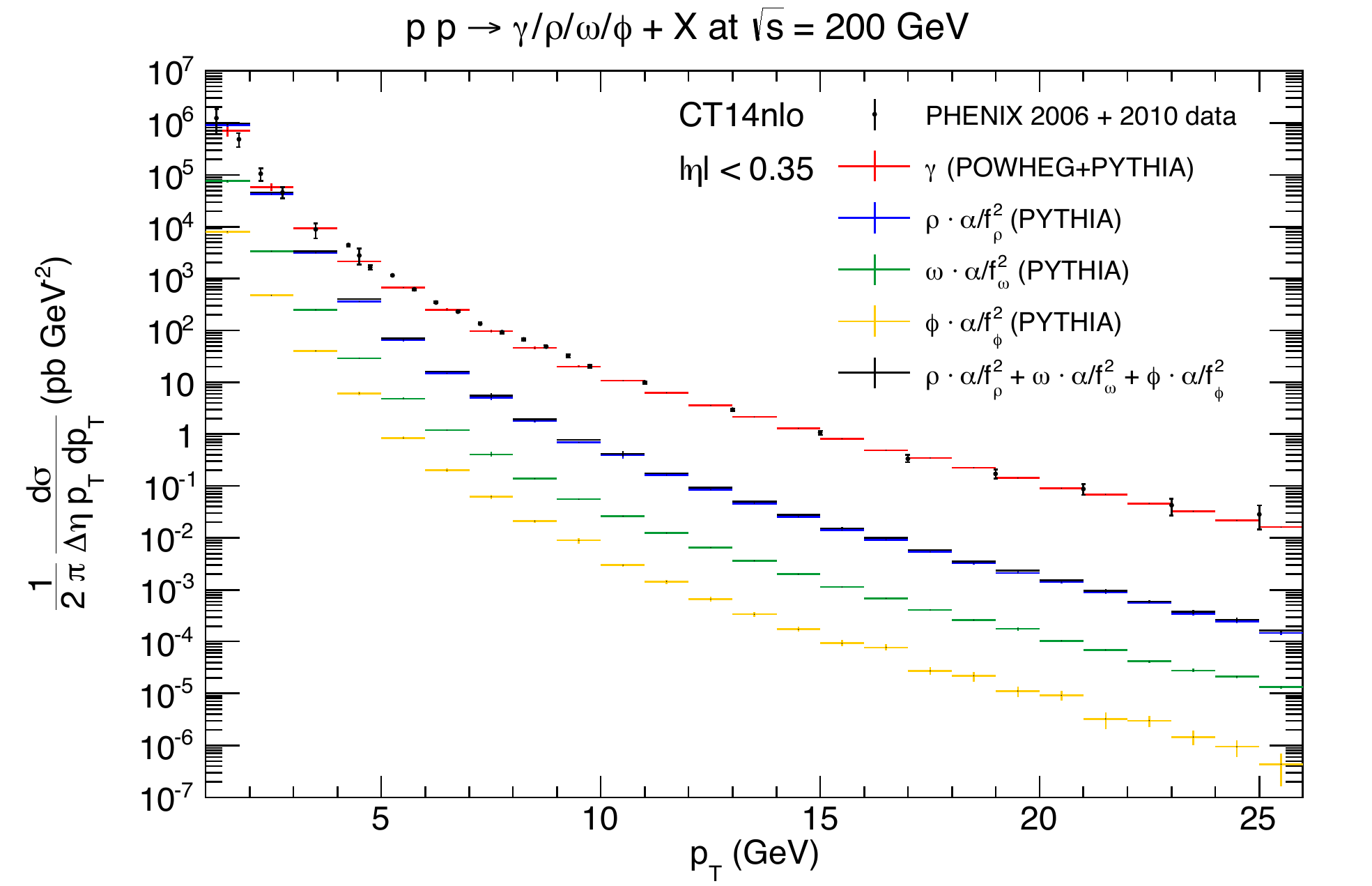}
 \caption{\label{fig:1b}Transverse-momentum spectra of pointlike photons and
 vector mesons fluctuating into photons in $pp$ collisions with $\sqrt{s}=200$ GeV
 at RHIC.}
\end{figure}
Strictly speaking, a description of fragmentation photons with partons showers
captures only the pointlike fragmentation component. As mentioned above,
this leads indeed to a correct description of the photon fragmentation function
at LEP \cite{Hoeche:2009xc,Bellm:2015jjp}. It is, however, well-known
that the photon fragmentation function also has a non-perturbative component
that is traditionally described with the Vector Meson Dominance (VMD) model
\bea
 |\gamma\rangle &=& \sum_{V=\rho,\omega,\phi}\frac{e}{f_V}|V\rangle,
% =
% \sqrt{\frac{e^2}{f_\rho^2}+\frac{e^2}{f_\omega^2}}(e_u^2+e_d^2)^{-1/2}
% (e_u|u\bar{u}\rangle+e_d|d\bar{d}\rangle)
% +\frac{e}{f_\phi}|s\bar{s}\rangle
% \label{eq:vmd}
\eea
where $f_V$ are the vector-meson decay constants \cite{Klasen:2002xb}.
We estimate the possible contributions from these long-range processes
in Fig.\ \ref{fig:1b}. As expected, the VMD contributions are dominated
by the lightest vector meson ($\rho$) and fall off more rapidly in $p_T$
than the pointlike photon contribution. Their contributions can only be
substantial at very low $p_T$. Like the vector mesons themselves, 
their fluctuations into photons might there indeed be sensitive to strong
medium effects,
contrary to the naive expectation that photons interact only electromagnetically.
A consistent combination of the pointlike and VMD contributions is
beyond the scope of this paper and is left for future work. It has, however,
previously been argued that in specific factorisation schemes such as the
DIS$_\gamma$ scheme the VMD component is completely negligible \cite{Gluck:1992zx}.
Since in this scheme an additional soft/collinear term $\ln[x^2(1-x)]$ is
resummed to all orders in the fragmentation function, similarly to the parton
shower, we do not expect a large VMD contribution in our NLO+PS approach.

\subsection{Fraction of isolated photons}

\begin{figure}[!h]
 \includegraphics[width=\linewidth]{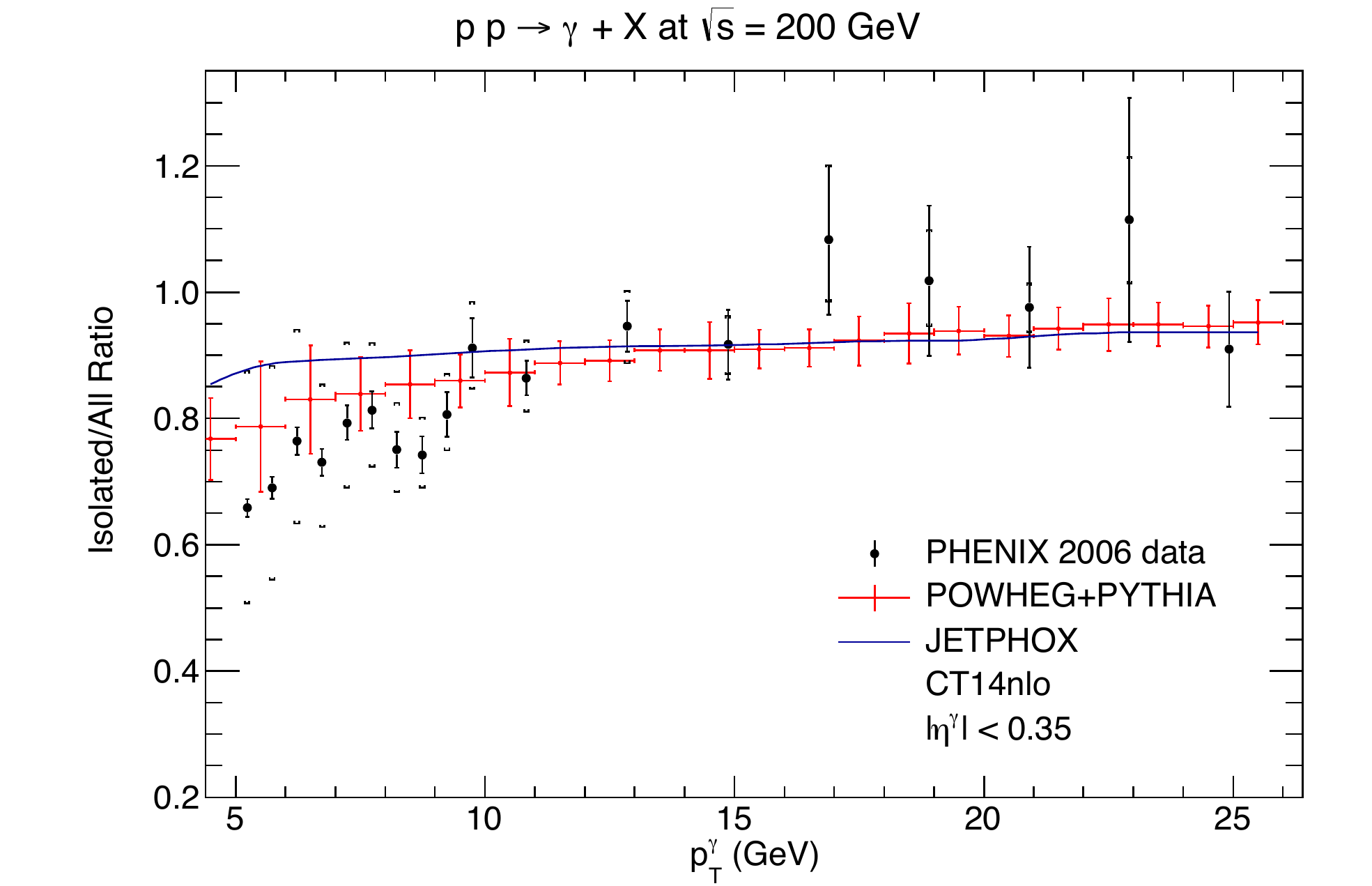}
 \caption{\label{fig:2}Ratio of isolated photons in NLO (blue) and
 NLO+PS (red) and compared to PHENIX data (black) \cite{Adare:2012yt}.}
\end{figure}

The fraction of isolated photons in the higher-$p_T$ PHENIX data set
is shown in Fig.\ \ref{fig:2} (black). The NLO calculation (blue) overestimates
%% data sets not used:
%% 0002 0075 0160 0195 0305 0314 0349
%
this fraction considerably in the low- and intermediate-$p_T$ region. This remains
true for all standard scale choices as already discussed in the experimental
publication \cite{Adare:2012yt}. There, the difference was temptatively
attributed to the underlying event activity or quark fragmentation contributions.
A comparison of the data with a LO+PS calculation from \PYTHIA{} did, however, not
show any drop in the low-$p_T$ region, either. As one can see in Fig.\ \ref{fig:2}, our
new NLO+PS calculation with \POWHEG+\PYTHIA{} (red) describes the PHENIX data
better, although the statistical error bars are still relatively
large. Note also that in this calculation the scale uncertainty cancels
completely, as the ratio is constructed from the same event sample in
the numerator and denominator and does not include any contributions
from the scale-dependent fragmentation function that affects the NLO
calculation differently in the numerator and denominator.

\subsection{Transverse-momentum spectrum of the associated charged hadron}

\begin{figure}[!h]
 \includegraphics[width=\linewidth]{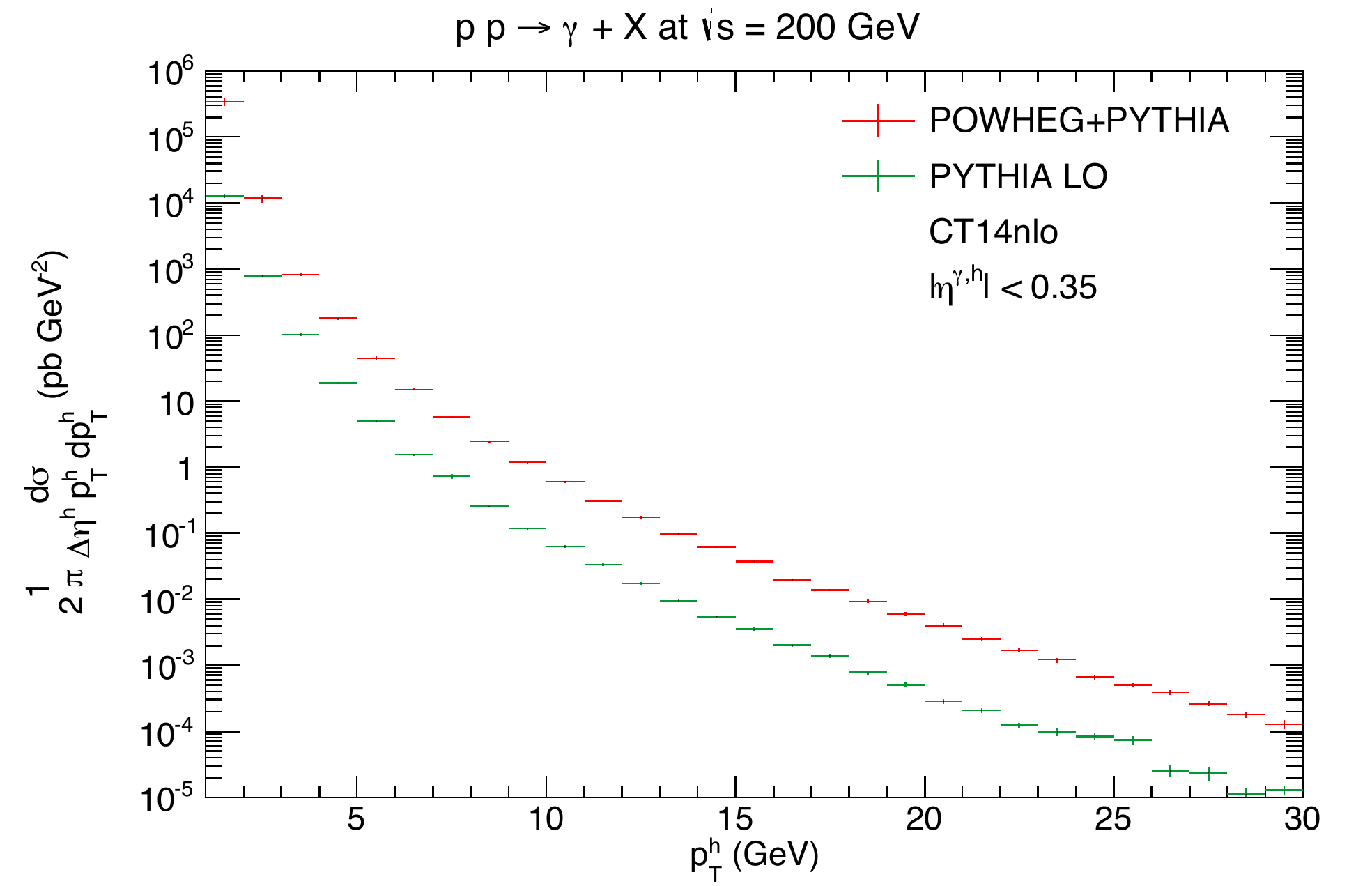}
 \caption{\label{fig:3}Transverse-momentum spectrum of the charged hadrons produced
 in association with a prompt photon in $pp$ collisions with $\sqrt{s}=200$ GeV
 at RHIC in LO+PS (green) and NLO+PS (red).}
\end{figure}

Important observables for the quark-gluon plasma are hadron energy loss
and jet quenching,
i.e.\ the energy loss of a hadronic jet induced by the hot medium.
The $p_T$-distribution of the charged hadrons produced in association with
the photon is therefore shown in Fig.\ \ref{fig:3}. Unfortunately,
%% data sets not used:
%% 0112 0379
%
it has not been measured in the cited PHENIX publications
\cite{Adare:2012vn,Adare:2012yt}. We therefore use the PHENIX
detector acceptance for charged hadrons $|\eta^{h}|<0.35$ quoted in the
preceding analysis of photon-hadron jet correlations (see below) \cite{Adare:2009vd}.
%and discard all events containing no charged hadron in the jet.
In LO, the leading jet transverse momentum equals that of the photon,
and indeed the LO+PS (green) and NLO+PS (red) distributions
follow those of the photon in Fig.\ \ref{fig:1} except for a shift
of the $p_T$-scale of roughly 20\% due to the missing neutral-hadron
contribution. In contrast to the
photon spectrum, however, the NLO $K$-factor remains almost constant
at high $p_T$ due to the fact that this region is very much QCD-like
in the sense that the observed charged hadron can also be balanced by other
partons rather than only a photon.

\subsection{Transverse-momentum balance of photons and charged hadrons}

Going one step further, one can also measure photon-hadron jet correlations
in $pp$ or heavy-ion collisions. They have the advantage that the 
photon (or more generally electroweak boson) balancing the charged-hadron
jet is not strongly influenced by the hot medium and can thus serve
as a gauge for the initial jet transverse momentum. An exact balance
holds, however, only at LO of perturbative QCD, so that deviations
can not be uniquely attributed to medium effects, but must also take
into account higher-order QCD corrections.

\begin{figure}[!h]
 \includegraphics[width=\linewidth]{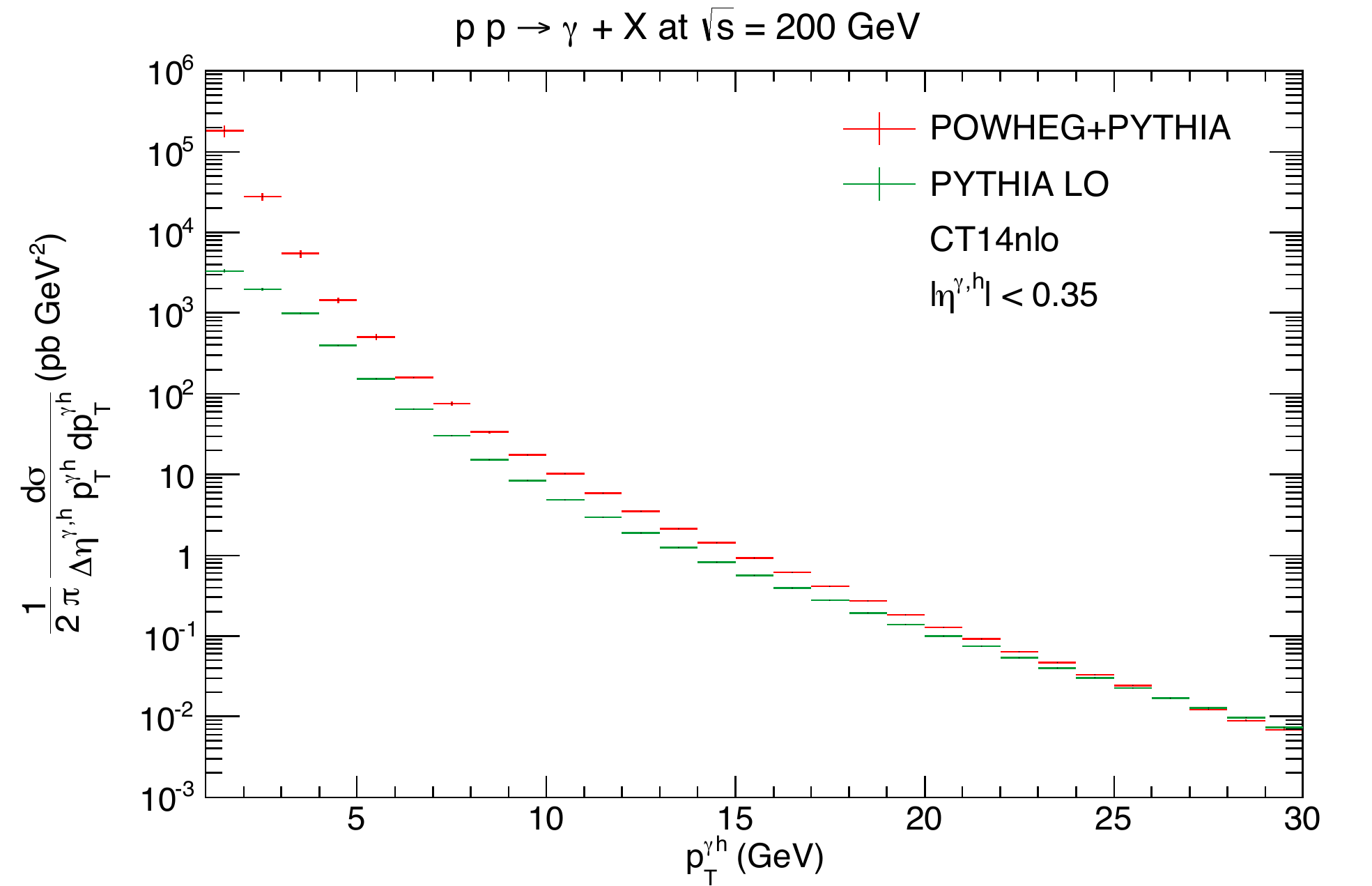}
 \caption{\label{fig:4}Transverse-momentum spectrum of the
 photon-hadron pair in $pp$ collisions with $\sqrt{s}=200$ GeV at RHIC
 in LO+PS (green) and NLO+PS (red).}
\end{figure}

In Fig.\ \ref{fig:4} we therefore show the distribution in the combined
%% data sets not used:
%% 0002 0075 0342 0349
%
photon-hadron transverse momentum $p_T^{\gamma h}$. To avoid the
non-perturbative region when all transverse momenta vanish,
we here apply individual cuts on $p_T^\gamma>1$ GeV and $p_T^h>1$
GeV.\footnote{In a realistic analysis, these two cuts should be
chosen unequal (see below) \cite{Klasen:1995xe}.}
 At LO (not shown), one then obtains a $\delta$-distribution
as the combined $p_T^{\gamma h}\to0$, while at NLO (also not shown)
the differential cross section is sensitive to the incomplete
cancellation of infrared divergences from the emission of one
additional soft parton. This region is resummed to all orders
by the parton shower in \PYTHIA{} already at LO (green), so that
this prediction exhibits a finite and physical turnover. The LO
normalisation is, however, still incorrect and modified by up
to two orders of magnitude in \POWHEG+\PYTHIA (red). Only at
this order can the $p_T$ imbalance be reliably predicted and this
observable subsequently applied to heavy-ion collisions in order
to extract genuine medium effects.

An important aspect of our NLO calculation is the appearance of
new partonic processes with no LO correspondence. In particular,
the process $q\bar{q}\to\gamma q\bar{q}$ with a recoiling quark
jet first enters at this order in addition to the process $q
\bar{q}\to\gamma gg$, which has just an additional gluon compared
to the LO process $q\bar{q}\to\gamma g$ with a recoiling gluon.
The recoiling quarks and gluons are then of course expected to behave
differently in the medium due to their different colour charges
and infrared behaviour. Another additional process first
entering at NLO is $gg\to\gamma q\bar{q}$ with no corresponding
process at LO. This gluon-initiated process is expected to be
more sensitive to the initial conditions of the heavy-ion collision,
in particular due to shadowing and saturation effects.

\subsection{Azimuthal correlation of photons and charged hadrons}

Besides jet quenching and the transverse-momentum imbalance, azimuthal
correlations of photons and hadron jets represent another important probe of
the quark-gluon plasma. They have therefore indeed been measured by
the PHENIX collaboration \cite{Adare:2009vd}, using unequal
transverse-momentum cuts on the photon ($p_T^\gamma\in[5;7]$ GeV) and
charged-hadron jet ($p_T^{h}\in[3;5]$ GeV) as required \cite{Klasen:1995xe}.
In particular the away-side correlation ($\Delta\phi\to\pi$)
has been found to be suppressed in 0-20\% central Au-Au collisions
for both decay and direct photons, which we can interpret as an
indication of decorrelation due to rescattering on the medium.

\begin{figure}[!h]
 \includegraphics[width=\linewidth]{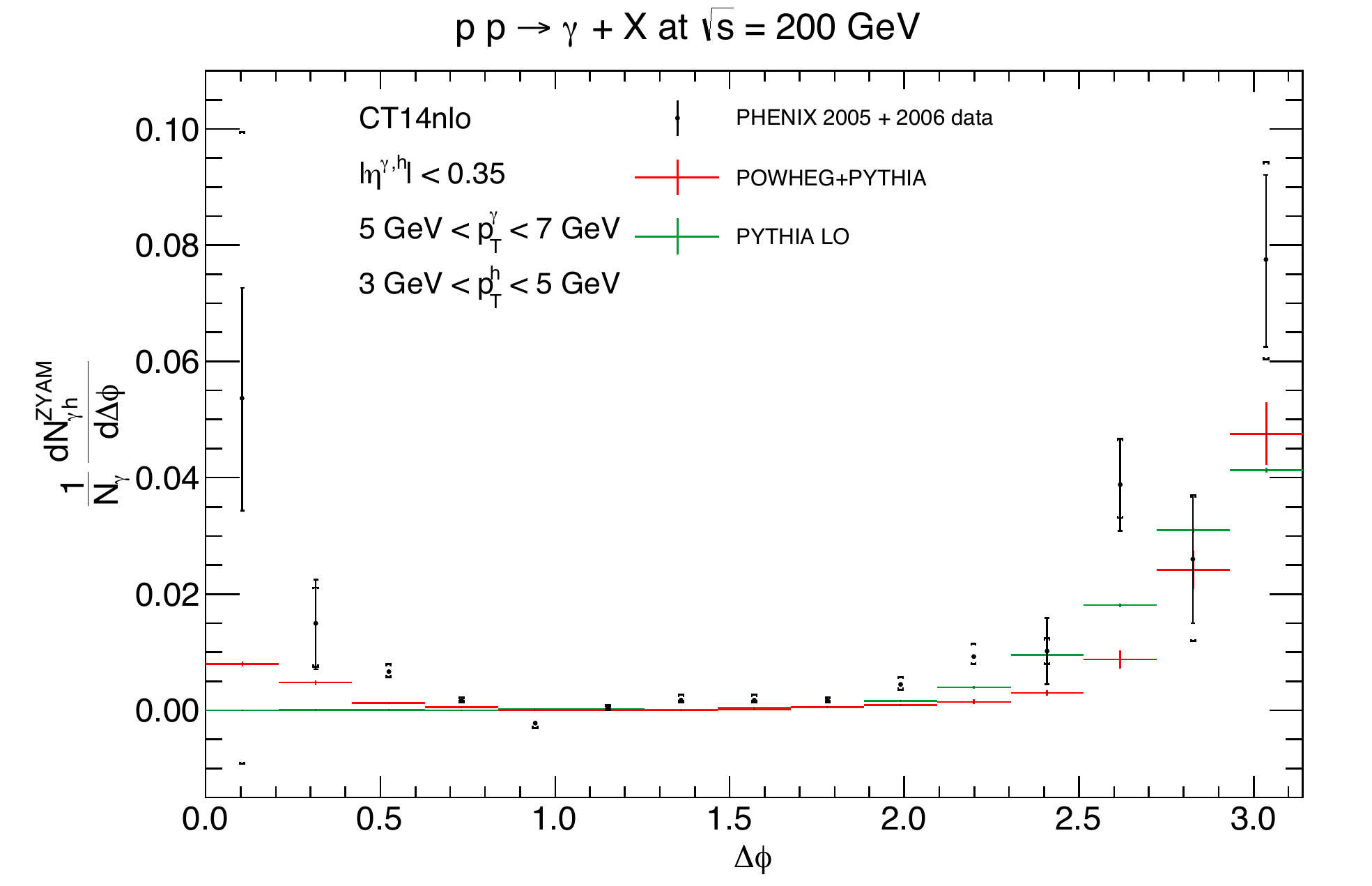}
 \caption{\label{fig:5}Azimuthal-angle correlation of the photon-hadron pair
 in $pp$ collisions with $\sqrt{s}=200$ GeV at RHIC in LO+PS (green) and
 NLO+PS (red) and compared to PHENIX data (black) \cite{Adare:2009vd}.}
\end{figure}

In Fig.\ \ref{fig:5} we reproduce the PHENIX data
%% data sets not used:
%% 0003 0008 0011 0020 0117 0130 0146 0186 0190 0193 0203 0218 0329 0339
%% 0389 0390 0392 0403 0410 0440 0448
%
(black) \cite{Adare:2009vd}, which have been obtained with a 
statistical subtraction of the decay photon background. In particular,
the cross section at the minimum of the correlation function has
been subtracted, assuming a Zero-Yield at Minimum (ZYAM),
and in addition it has been normalised to the total number of 
trigger photons. Both the LO and NLO calculations predict unphysical
results for this observable (a $\delta$-distribution at $\pi$
and vanishing results for jets with $\Delta\phi$ below $\pi/2$
\cite{Saimpert:2015oka}) and are therefore not shown.
In contrast, our NLO+PS calculation with \POWHEG+\PYTHIA{} (red) exhibits
a physical behavior with finite results in both limits $\Delta\phi\to\{0;\pi\}$
and describes the PHENIX data quite well, while the LO+PS prediction with
\PYTHIA{} alone (green) cannot correctly describe the region 
$\Delta\phi\to0$ (see below) and also has a wrong normalisation (not shown).

\begin{figure}[!h]
 \includegraphics[width=\linewidth]{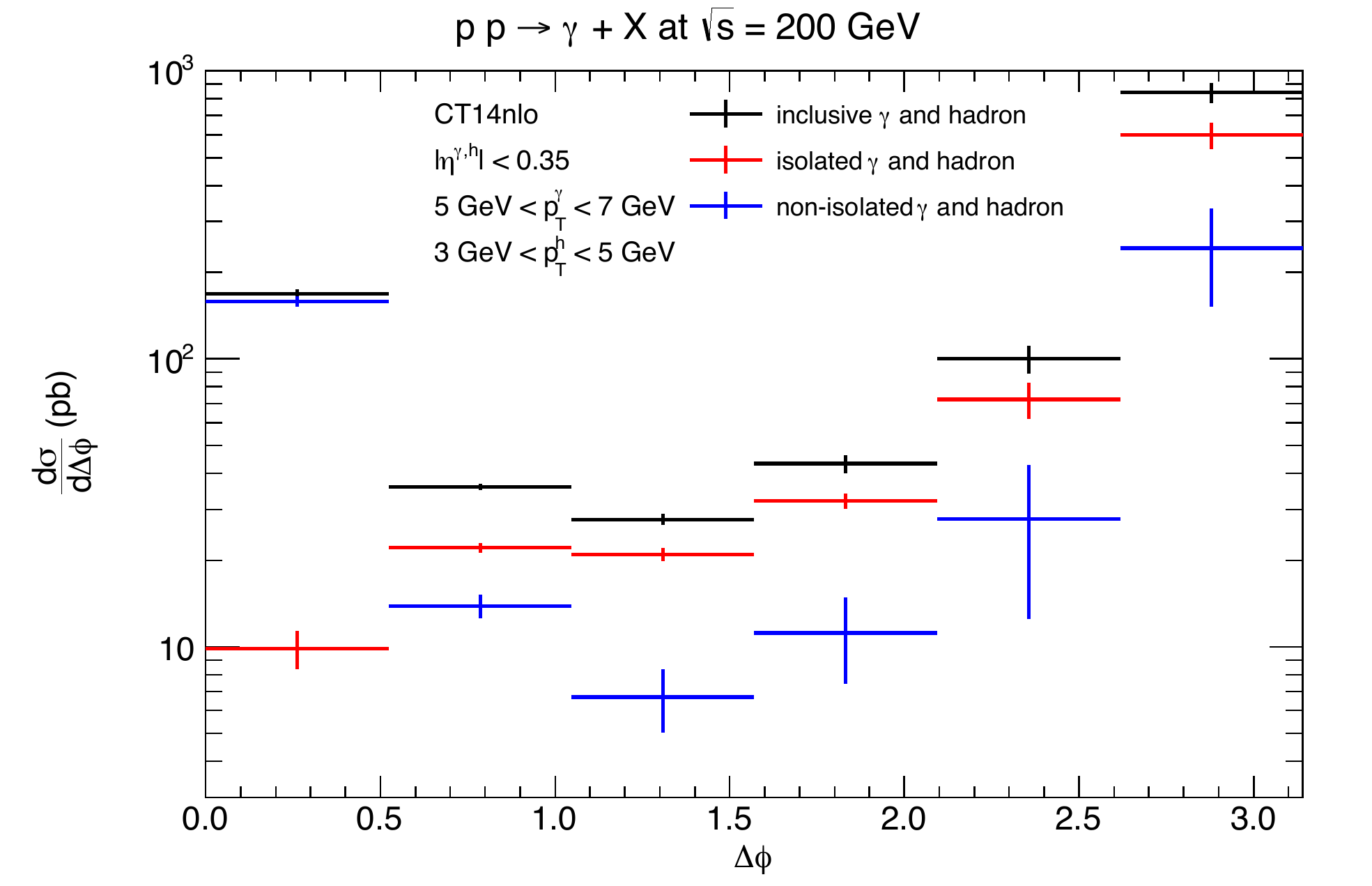}
 \caption{\label{fig:6}Isolated (red), non-isolated (blue) and total (black) photon
 contributions to the azimuthal-angle correlation of the photon-hadron pair in $pp$
 collisions with $\sqrt{s}=200$ GeV at RHIC.}
\end{figure}

To better understand the individual contributions to the near- and away-side
correlation function, we show in Fig.\ \ref{fig:6} the isolated (red),
%% data sets: same as previous figure
%
non-isolated (blue) and total (black) photon contributions to the azimuthal-angle
correlation of the photon-hadron pair. As expected, the non-isolated photons originating
from fragmentation processes dominate the cross section in the near-side region, i.e.\
they are mostly collinear to the parton fragmenting into the observed hadron ($\Delta\phi\simeq0$), and these NLO
processes are of course missing in the LO+PS calculation. At the level of 35\%,
they also contribute in the away-side region ($\Delta\phi\simeq\pi$), where they
originate from fragmentation from the (unobserved) second parton. This region is,
however, dominated by back-to-back photons and partons as in LO, so that \PYTHIA{}
captures the essence of the physics in this region. Note that our predictions in
Fig.\ \ref{fig:6} are neither subtracted to ZYAM nor normalised for possible
future comparisons of absolute cross sections. Remember also
that in heavy-ion collisions isolation cuts can not be used due to the
large hadronic underlying event in the low-$p_T$ region, where one wants
to extract the thermal-photon spectrum, so that a correct description of
fragmentation processes is very important.

\subsection{Pseudorapidity correlation of photons and charged hadrons}

\begin{figure}[!h]
 \includegraphics[width=\linewidth]{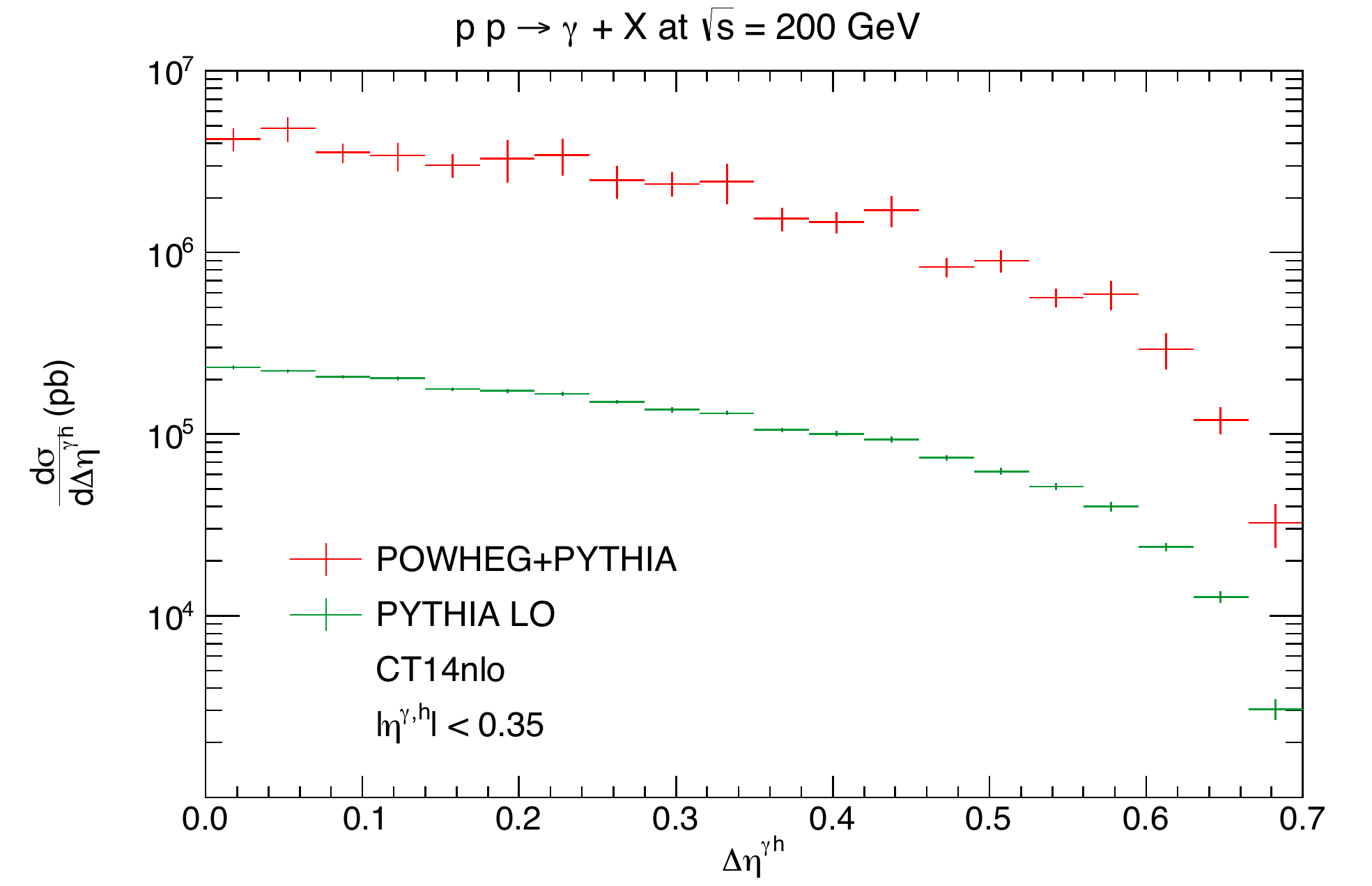}
 \caption{\label{fig:7}Pseudorapidity correlation of the photon-hadron pair
 in $pp$ collisions with $\sqrt{s}=200$ GeV at RHIC in LO+PS (green) and
 NLO+PS (red).}
\end{figure}

For completeness, we show in Fig.\ \ref{fig:7} the correlation in
%% data sets not used:
%% 0021 0025 0032 0075 0113 0144 0163 0169 0179 0181 0275 0277 0289 0286
%% 0314 0336 0343
%
radidity of the photon-hadron pair with the same individual pseudorapidity cuts
as in the previous subsection and $p_T^{\gamma},p_T^{h}>1$ GeV.
While the pseudorapidity range accessible
to the RHIC detectors has so far been quite limited, future upgrades at RHIC
or measurements at the LHC bear important potential for studies of the
low-$x$ region and therefore of the initial conditions of the formation
of the QGP e.g.\ from a colour-glass condensate.

%%%%%%%%%%%%%% Begin Section 5 %%%%%%%%%%%%%%%%%%%%%%%%%%%%%%%%%%%%%%%%%
\section{Conclusion}
\label{sec:5}

In this paper we have presented a calculation of direct photon
production at NLO QCD, a matching of this calculation with parton
showers using \POWHEGBOX{}, and a detailed phenomenological analysis
of PHENIX data on prompt photon production and photon-hadron jet
correlations in $pp$ collisions at RHIC energies.
Our work was motivated by the
facts that the inclusion of photons in parton showers is highly
non-trivial, that prompt photons are important probes of the QGP in
heavy-ion collisions, for which $pp$ form the indispensable baseline,
and that they give furthermore access to photon-hadron jet correlations and
studies of jet quenching.
To this end, we have described in detail the analytical and numerical
validations of our calculations at different stages and then
solutions to various encountered difficulties such as the suppression of
divergent Born contributions, the symmetrisation of parton splittings
involving photons, and the enhancement of QED radiation. 

The application of our NLO+PS calculations to PHENIX data taken at RHIC
has led to important improvements compared to both LO+PS and pure NLO:
in the description of the low-$p_T$ inclusive photon spectrum, of the
fraction of isolated photons contained in this data sample, and of the
azimuthal correlations of photons and charged hadrons. In addition, we have made
predictions for the $p_T$ spectra of the associated charged hadron and for the
$p_T$ balance of the photon-hadron pair as well as their pseudorapidity correlation
and have decomposed the azimuthal correlation function into
fragmentation and non-fragmentation components. 

As the next step, our calculation can easily be applied to $pA$ and
$AA$ collisions for studies of cold nuclear effects at RHIC or the LHC.
Subsequently, one can tackle the implementation of rescattering on
the medium along the lines of Ref.\ \cite{KunnawalkamElayavalli:2016ttl},
but including NLO corrections. A consistent combination of pointlike and
VMD contributions to photon fragmentation in Monte Carlo generators is
also envisaged in future work.

%%%%%%%%%%%%%% Begin Acknowledgment %%%%%%%%%%%%%%%%%%%%%%%%%%%%%%%%%%%%
\section*{Acknowledgments}

We thank C.\ Klein-B\"osing for useful discussions. This work has been
supported by the BMBF under contract 05H15PMCCA.

%%%%%%%%%%%%%% Begin Bibliography %%%%%%%%%%%%%%%%%%%%%%%%%%%%%%%%%%%%%%
\bibliographystyle{JHEP}

\bibliography{paper}

\begin{thebibliography}{}

%\cite{Adcox:2004mh}
\bibitem{Adcox:2004mh}
  K.~Adcox {\it et al.} [PHENIX Collaboration],
  %``Formation of dense partonic matter in relativistic nucleus-nucleus collisions at RHIC: Experimental evaluation by the PHENIX collaboration,''
  Nucl.\ Phys.\ A {\bf 757} (2005) 184
  doi:10.1016/j.nuclphysa.2005.03.086
  [nucl-ex/0410003].
  %%CITATION = doi:10.1016/j.nuclphysa.2005.03.086;%%
  %2210 citations counted in INSPIRE as of 22 Aug 2016

%\cite{Adams:2005dq}
\bibitem{Adams:2005dq}
  J.~Adams {\it et al.} [STAR Collaboration],
  %``Experimental and theoretical challenges in the search for the quark gluon plasma: The STAR Collaboration's critical assessment of the evidence from RHIC collisions,''
  Nucl.\ Phys.\ A {\bf 757} (2005) 102
  doi:10.1016/j.nuclphysa.2005.03.085
  [nucl-ex/0501009].
  %%CITATION = doi:10.1016/j.nuclphysa.2005.03.085;%%
  %2354 citations counted in INSPIRE as of 22 Aug 2016

%\cite{Aamodt:2010pa}
\bibitem{Aamodt:2010pa}
  K.~Aamodt {\it et al.} [ALICE Collaboration],
  %``Elliptic flow of charged particles in Pb-Pb collisions at 2.76 TeV,''
  Phys.\ Rev.\ Lett.\  {\bf 105} (2010) 252302
  doi:10.1103/PhysRevLett.105.252302
  [arXiv:1011.3914 [nucl-ex]].
  %%CITATION = doi:10.1103/PhysRevLett.105.252302;%%
  %599 citations counted in INSPIRE as of 22 Aug 2016

%\cite{Aad:2010bu}
\bibitem{Aad:2010bu}
  G.~Aad {\it et al.} [ATLAS Collaboration],
  %``Observation of a Centrality-Dependent Dijet Asymmetry in Lead-Lead Collisions at $\sqrt{s_{NN}}=2.77$ TeV with the ATLAS Detector at the LHC,''
  Phys.\ Rev.\ Lett.\  {\bf 105} (2010) 252303
  doi:10.1103/PhysRevLett.105.252303
  [arXiv:1011.6182 [hep-ex]].
  %%CITATION = doi:10.1103/PhysRevLett.105.252303;%%
  %549 citations counted in INSPIRE as of 22 Aug 2016

%\cite{Chatrchyan:2011sx}
\bibitem{Chatrchyan:2011sx}
  S.~Chatrchyan {\it et al.} [CMS Collaboration],
  %``Observation and studies of jet quenching in PbPb collisions at nucleon-nucleon center-of-mass energy = 2.76 TeV,''
  Phys.\ Rev.\ C {\bf 84} (2011) 024906
  doi:10.1103/PhysRevC.84.024906
  [arXiv:1102.1957 [nucl-ex]].
  %%CITATION = doi:10.1103/PhysRevC.84.024906;%%
  %500 citations counted in INSPIRE as of 22 Aug 2016

%\cite{Wilde:2012wc}
\bibitem{Wilde:2012wc}
  M.~Wilde [ALICE Collaboration],
  %``Measurement of Direct Photons in pp and Pb-Pb Collisions with ALICE,''
  Nucl.\ Phys.\ A904-905 {\bf 2013} (2013) 573c
  [arXiv:1210.5958 [hep-ex]].
  %%CITATION = ARXIV:1210.5958;%%
  %12 citations counted in INSPIRE as of 01 Jul 2013

%\cite{Klasen:2013mga}
\bibitem{Klasen:2013mga}
  M.~Klasen, C.~Klein-B\"osing, F.~K\"onig and J.~P.~Wessels,
  %``How robust is a thermal photon interpretation of the ALICE low-p_T data?,''
  JHEP {\bf 1310} (2013) 119
  doi:10.1007/JHEP10(2013)119
  [arXiv:1307.7034 [hep-ph]].
  %%CITATION = doi:10.1007/JHEP10(2013)119;%%
  %15 citations counted in INSPIRE as of 25 Mar 2016

%\cite{Klasen:2014xra}
\bibitem{Klasen:2014xra}
  M.~Klasen, F.~K\"onig, C.~Klein-B\"osing and J.~P.~Wessels,
  %``QCD analysis and effective temperature of direct photons in lead-lead collisions at the LHC,''
  Nucl.\ Part.\ Phys.\ Proc.\  {\bf 273-275} (2016) 1509
  doi:10.1016/j.nuclphysbps.2015.09.244
  [arXiv:1409.3363 [hep-ph]].
  %%CITATION = doi:10.1016/j.nuclphysbps.2015.09.244;%%

%\cite{Adam:2015lda}
\bibitem{Adam:2015lda}
  J.~Adam {\it et al.} [ALICE Collaboration],
  %``Direct photon production in Pb-Pb collisions at $\sqrt{s_\rm{NN}} =$ 2.76 TeV,''
  Phys.\ Lett.\ B {\bf 754} (2016) 235
  doi:10.1016/j.physletb.2016.01.020
  [arXiv:1509.07324 [nucl-ex]].
  %%CITATION = doi:10.1016/j.physletb.2016.01.020;%%
  %23 citations counted in INSPIRE as of 23 Aug 2016

%\cite{Adare:2008ab}
\bibitem{Adare:2008ab}
  A.~Adare {\it et al.} [PHENIX Collaboration],
  %``Enhanced production of direct photons in Au+Au collisions at $\sqrt{s_{NN}}=200$ GeV and implications for the initial temperature,''
  Phys.\ Rev.\ Lett.\  {\bf 104} (2010) 132301
  doi:10.1103/PhysRevLett.104.132301
  [arXiv:0804.4168 [nucl-ex]].
  %%CITATION = doi:10.1103/PhysRevLett.104.132301;%%
  %391 citations counted in INSPIRE as of 23 Aug 2016

%\cite{Adare:2009vd}
\bibitem{Adare:2009vd}
  A.~Adare {\it et al.} [PHENIX Collaboration],
  %``Photon-Hadron Jet Correlations in p+p and Au+Au Collisions at s**(1/2) = 200-GeV,''
  Phys.\ Rev.\ C {\bf 80} (2009) 024908
  doi:10.1103/PhysRevC.80.024908
  [arXiv:0903.3399 [nucl-ex]].
  %%CITATION = doi:10.1103/PhysRevC.80.024908;%%
  %67 citations counted in INSPIRE as of 25 Mar 2016

%\cite{Arbor:2013ola}
\bibitem{Arbor:2013ola}
  N.~Arbor [ALICE Collaboration],
  %``Recent photon physics results from the ALICE experiment at the LHC,''
  EPJ Web Conf.\  {\bf 60} (2013) 13011.
  doi:10.1051/epjconf/20136013011
  %%CITATION = doi:10.1051/epjconf/20136013011;%%
  %1 citations counted in INSPIRE as of 25 Mar 2016

%\cite{Klasen:2002xb}
\bibitem{Klasen:2002xb}
  M.~Klasen,
  %``Theory of hard photoproduction,''
  Rev.\ Mod.\ Phys.\  {\bf 74} (2002) 1221
  doi:10.1103/RevModPhys.74.1221
  [hep-ph/0206169].
  %%CITATION = doi:10.1103/RevModPhys.74.1221;%%
  %89 citations counted in INSPIRE as of 23 Aug 2016

%\cite{Aurenche:2006vj}
\bibitem{Aurenche:2006vj}
  P.~Aurenche, M.~Fontannaz, J.~-P.~Guillet, E.~Pilon and M.~Werlen,
  %``A New critical study of photon production in hadronic collisions,''
  Phys.\ Rev.\ D {\bf 73} (2006) 094007
  [hep-ph/0602133].
  %%CITATION = HEP-PH/0602133;%%
  %124 citations counted in INSPIRE as of 01 Jul 2013

%\cite{Klasen:2014xfa}
\bibitem{Klasen:2014xfa}
  M.~Klasen and F.~K\"onig,
  %``New information on photon fragmentation functions,''
  Eur.\ Phys.\ J.\ C {\bf 74} (2014) no.8,  3009
  doi:10.1140/epjc/s10052-014-3009-x
  [arXiv:1403.2290 [hep-ph]].
  %%CITATION = doi:10.1140/epjc/s10052-014-3009-x;%%
  %3 citations counted in INSPIRE as of 25 Mar 2016

%\cite{Frixione:1998jh}
\bibitem{Frixione:1998jh}
  S.~Frixione,
  %``Isolated photons in perturbative QCD,''
  Phys.\ Lett.\ B {\bf 429} (1998) 369
  doi:10.1016/S0370-2693(98)00454-7
  [hep-ph/9801442].
  %%CITATION = doi:10.1016/S0370-2693(98)00454-7;%%
  %203 citations counted in INSPIRE as of 23 Aug 2016

%\cite{Berger:1998ev}
\bibitem{Berger:1998ev}
  E.~L.~Berger, L.~E.~Gordon and M.~Klasen,
  %``Massive lepton pairs as a prompt photon surrogate,''
  Phys.\ Rev.\ D {\bf 58} (1998) 074012
  doi:10.1103/PhysRevD.58.074012
  [hep-ph/9803387].
  %%CITATION = doi:10.1103/PhysRevD.58.074012;%%
  %58 citations counted in INSPIRE as of 23 Aug 2016

%\cite{Berger:1999es}
\bibitem{Berger:1999es}
  E.~L.~Berger, L.~E.~Gordon and M.~Klasen,
  %``Spin dependence of massive lepton pair production in proton proton collisions,''
  Phys.\ Rev.\ D {\bf 62} (2000) 014014
  doi:10.1103/PhysRevD.62.014014
  [hep-ph/9909446].
  %%CITATION = doi:10.1103/PhysRevD.62.014014;%%
  %20 citations counted in INSPIRE as of 23 Aug 2016

%\cite{Brandt:2013hoa}
\bibitem{Brandt:2013hoa}
  M.~Klasen and M.~Brandt,
  %``Parton densities from LHC vector boson production at small and large transverse momenta,''
  Phys.\ Rev.\ D {\bf 88} (2013) 054002
  doi:10.1103/PhysRevD.88.054002
  [arXiv:1305.5677 [hep-ph]].
  %%CITATION = doi:10.1103/PhysRevD.88.054002;%%
  %13 citations counted in INSPIRE as of 23 Aug 2016

%\cite{Brandt:2014vva}
\bibitem{Brandt:2014vva}
  M.~Brandt, M.~Klasen and F.~K\"onig,
  %``Nuclear parton density modifications from low-mass lepton pair production at the LHC,''
  Nucl.\ Phys.\ A {\bf 927} (2014) 78
  doi:10.1016/j.nuclphysa.2014.03.024
  [arXiv:1401.6817 [hep-ph]].
  %%CITATION = doi:10.1016/j.nuclphysa.2014.03.024;%%
  %6 citations counted in INSPIRE as of 23 Aug 2016

%\cite{Hoeche:2009xc}
\bibitem{Hoeche:2009xc}
  S.~H\"oche, S.~Schumann and F.~Siegert,
  %``Hard photon production and matrix-element parton-shower merging,''
  Phys.\ Rev.\ D {\bf 81} (2010) 034026
  doi:10.1103/PhysRevD.81.034026
  [arXiv:0912.3501 [hep-ph]].
  %%CITATION = doi:10.1103/PhysRevD.81.034026;%%
  %80 citations counted in INSPIRE as of 25 Mar 2016

%\cite{Bellm:2015jjp}
\bibitem{Bellm:2015jjp}
  J.~Bellm {\it et al.},
  %``Herwig 7.0/Herwig++ 3.0 release note,''
  Eur.\ Phys.\ J.\ C {\bf 76} (2016) no.4,  196
  doi:10.1140/epjc/s10052-016-4018-8
  [arXiv:1512.01178 [hep-ph]].
  %%CITATION = doi:10.1140/epjc/s10052-016-4018-8;%%
  %33 citations counted in INSPIRE as of 23 Aug 2016

%\cite{Frixione:2002ik}
\bibitem{Frixione:2002ik}
  S.~Frixione and B.~R.~Webber,
  %``Matching NLO QCD computations and parton shower simulations,''
  JHEP {\bf 0206} (2002) 029
  doi:10.1088/1126-6708/2002/06/029
  [hep-ph/0204244].
  %%CITATION = doi:10.1088/1126-6708/2002/06/029;%%
  %2162 citations counted in INSPIRE as of 23 Aug 2016

%\cite{Frixione:2007vw}
\bibitem{Frixione:2007vw}
  S.~Frixione, P.~Nason and C.~Oleari,
  %``Matching NLO QCD computations with Parton Shower simulations: the POWHEG method,''
  JHEP {\bf 0711} (2007) 070
  doi:10.1088/1126-6708/2007/11/070
  [arXiv:0709.2092 [hep-ph]].
  %%CITATION = doi:10.1088/1126-6708/2007/11/070;%%
  %1507 citations counted in INSPIRE as of 31 mars 2016

%\cite{D'Errico:2011sd}
\bibitem{D'Errico:2011sd}
  L.~D'Errico and P.~Richardson,
  %``Next-to-Leading-Order Monte Carlo Simulation of Diphoton Production in Hadronic Collisions,''
  JHEP {\bf 1202} (2012) 130
  doi:10.1007/JHEP02(2012)130
  [arXiv:1106.3939 [hep-ph]].
  %%CITATION = doi:10.1007/JHEP02(2012)130;%%
  %20 citations counted in INSPIRE as of 25 Mar 2016

%\cite{Barze:2012tt}
\bibitem{Barze:2012tt}
  L.~Barze, G.~Montagna, P.~Nason, O.~Nicrosini and F.~Piccinini,
  %``Implementation of electroweak corrections in the POWHEG BOX: single W production,''
  JHEP {\bf 1204} (2012) 037
  doi:10.1007/JHEP04(2012)037
  [arXiv:1202.0465 [hep-ph]].
  %%CITATION = doi:10.1007/JHEP04(2012)037;%%
  %35 citations counted in INSPIRE as of 31 mars 2016

%\cite{Barze:2014zba}
\bibitem{Barze:2014zba}
  L.~Barze, M.~Chiesa, G.~Montagna, P.~Nason, O.~Nicrosini, F.~Piccinini and V.~Prosperi,
  %``W$\gamma$ production in hadronic collisions using the POWHEG+MiNLO method,''
  JHEP {\bf 1412} (2014) 039
  doi:10.1007/JHEP12(2014)039
  [arXiv:1408.5766 [hep-ph]].
  %%CITATION = doi:10.1007/JHEP12(2014)039;%%
  %8 citations counted in INSPIRE as of 23 Aug 2016

%\cite{Alioli:2010xd}
\bibitem{Alioli:2010xd}
  S.~Alioli, P.~Nason, C.~Oleari and E.~Re,
  %``A general framework for implementing NLO calculations in shower Monte Carlo programs: the POWHEG BOX,''
  JHEP {\bf 1006} (2010) 043
  doi:10.1007/JHEP06(2010)043
  [arXiv:1002.2581 [hep-ph]].
  %%CITATION = doi:10.1007/JHEP06(2010)043;%%
  %1045 citations counted in INSPIRE as of 31 mars 2016

%\cite{Hahn:1998yk}
\bibitem{Hahn:1998yk}
  T.~Hahn and M.~Perez-Victoria,
  %``Automatized one loop calculations in four-dimensions and D-dimensions,''
  Comput.\ Phys.\ Commun.\  {\bf 118} (1999) 153
  doi:10.1016/S0010-4655(98)00173-8
  [hep-ph/9807565].
  %%CITATION = doi:10.1016/S0010-4655(98)00173-8;%%
  %1111 citations counted in INSPIRE as of 25 mars 2016

%\cite{Alwall:2014hca}
\bibitem{Alwall:2014hca}
  J.~Alwall {\it et al.},
  %``The automated computation of tree-level and next-to-leading order differential cross sections, and their matching to parton shower simulations,''
  JHEP {\bf 1407} (2014) 079
  doi:10.1007/JHEP07(2014)079
  [arXiv:1405.0301 [hep-ph]].
  %%CITATION = doi:10.1007/JHEP07(2014)079;%%
  %856 citations counted in INSPIRE as of 31 mars 2016

%\cite{Berger:1983yi}
\bibitem{Berger:1983yi}
  E.~L.~Berger, E.~Braaten and R.~D.~Field,
  %``Large p(T) Production of Single and Double Photons in Proton Proton and Pion-Proton Collisions,''
  Nucl.\ Phys.\ B {\bf 239} (1984) 52.
  doi:10.1016/0550-3213(84)90084-1
  %%CITATION = doi:10.1016/0550-3213(84)90084-1;%%
  %150 citations counted in INSPIRE as of 24 Aug 2016

%\cite{Owens:1986mp}
\bibitem{Owens:1986mp}
  J.~F.~Owens,
  %``Large Momentum Transfer Production of Direct Photons, Jets, and Particles,''
  Rev.\ Mod.\ Phys.\  {\bf 59} (1987) 465.
  doi:10.1103/RevModPhys.59.465
  %%CITATION = doi:10.1103/RevModPhys.59.465;%%
  %556 citations counted in INSPIRE as of 24 Aug 2016

%\cite{Aurenche:1987fs}
\bibitem{Aurenche:1987fs}
  P.~Aurenche, R.~Baier, M.~Fontannaz and D.~Schiff,
  %``Prompt Photon Production at Large p(T) Scheme Invariant QCD Predictions and Comparison with Experiment,''
  Nucl.\ Phys.\ B {\bf 297} (1988) 661.
  doi:10.1016/0550-3213(88)90553-6
  %%CITATION = doi:10.1016/0550-3213(88)90553-6;%%
  %299 citations counted in INSPIRE as of 24 Aug 2016

%\cite{Gordon:1993qc}
\bibitem{Gordon:1993qc}
  L.~E.~Gordon and W.~Vogelsang,
  %``Polarized and unpolarized prompt photon production beyond the leading order,''
  Phys.\ Rev.\ D {\bf 48} (1993) 3136.
  doi:10.1103/PhysRevD.48.3136
  %%CITATION = doi:10.1103/PhysRevD.48.3136;%%
  %232 citations counted in INSPIRE as of 24 Aug 2016

%\cite{Vermaseren:2000nd}
\bibitem{Vermaseren:2000nd}
  J.~A.~M.~Vermaseren,
  %``New features of FORM,''
  math-ph/0010025.
  %%CITATION = MATH-PH/0010025;%%
  %1145 citations counted in INSPIRE as of 25 mars 2016

%\cite{Passarino:1978jh}
\bibitem{Passarino:1978jh}
  G.~Passarino and M.~J.~G.~Veltman,
  %``One Loop Corrections for e+ e- Annihilation Into mu+ mu- in the Weinberg Model,''
  Nucl.\ Phys.\ B {\bf 160} (1979) 151.
  doi:10.1016/0550-3213(79)90234-7
  %%CITATION = doi:10.1016/0550-3213(79)90234-7;%%
  %1970 citations counted in INSPIRE as of 25 Mar 2016

%\cite{Bardeen:1978yd}
\bibitem{Bardeen:1978yd}
  W.~A.~Bardeen, A.~J.~Buras, D.~W.~Duke and T.~Muta,
  %``Deep Inelastic Scattering Beyond the Leading Order in Asymptotically Free Gauge Theories,''
  Phys.\ Rev.\ D {\bf 18} (1978) 3998.
  doi:10.1103/PhysRevD.18.3998
  %%CITATION = doi:10.1103/PhysRevD.18.3998;%%
  %1448 citations counted in INSPIRE as of 24 Aug 2016

%\cite{Catani:1996vz}
\bibitem{Catani:1996vz}
  S.~Catani and M.~H.~Seymour,
  %``A General algorithm for calculating jet cross-sections in NLO QCD,''
  Nucl.\ Phys.\ B {\bf 485} (1997) 291
   Erratum: [Nucl.\ Phys.\ B {\bf 510} (1998) 503]
  doi:10.1016/S0550-3213(96)00589-5
  [hep-ph/9605323].
  %%CITATION = doi:10.1016/S0550-3213(96)00589-5;%%
  %1301 citations counted in INSPIRE as of 31 mars 2016

%\cite{Hasegawa:2009tx}
\bibitem{Hasegawa:2009tx}
  K.~Hasegawa, S.~Moch and P.~Uwer,
  %``AutoDipole: Automated generation of dipole subtraction terms,''
  Comput.\ Phys.\ Commun.\  {\bf 181} (2010) 1802
  doi:10.1016/j.cpc.2010.06.044
  [arXiv:0911.4371 [hep-ph]].
  %%CITATION = doi:10.1016/j.cpc.2010.06.044;%%
  %48 citations counted in INSPIRE as of 24 Aug 2016

%\cite{Sjostrand:2014zea}
\bibitem{Sjostrand:2014zea}
  T.~Sj\"ostrand {\it et al.},
  %``An Introduction to PYTHIA 8.2,''
  Comput.\ Phys.\ Commun.\  {\bf 191} (2015) 159
  doi:10.1016/j.cpc.2015.01.024
  [arXiv:1410.3012 [hep-ph]].
  %%CITATION = doi:10.1016/j.cpc.2015.01.024;%%
  %181 citations counted in INSPIRE as of 31 mars 2016

%\cite{Frixione:1995ms}
\bibitem{Frixione:1995ms}
  S.~Frixione, Z.~Kunszt and A.~Signer,
  %``Three jet cross-sections to next-to-leading order,''
  Nucl.\ Phys.\ B {\bf 467} (1996) 399
  doi:10.1016/0550-3213(96)00110-1
  [hep-ph/9512328].
  %%CITATION = doi:10.1016/0550-3213(96)00110-1;%%
  %564 citations counted in INSPIRE as of 24 Aug 2016

%\cite{Alioli:2010xa}
\bibitem{Alioli:2010xa}
  S.~Alioli, K.~Hamilton, P.~Nason, C.~Oleari and E.~Re,
  %``Jet pair production in POWHEG,''
  JHEP {\bf 1104} (2011) 081
  doi:10.1007/JHEP04(2011)081
  [arXiv:1012.3380 [hep-ph]].
  %%CITATION = doi:10.1007/JHEP04(2011)081;%%
  %157 citations counted in INSPIRE as of 25 Aug 2016

%\cite{Nason:2013uba}
\bibitem{Nason:2013uba}
  P.~Nason and C.~Oleari,
  %``Generation cuts and Born suppression in POWHEG,''
  arXiv:1303.3922 [hep-ph].
  %%CITATION = ARXIV:1303.3922;%%
  %15 citations counted in INSPIRE as of 25 Aug 2016

%\cite{Lonnblad:2012hz}
\bibitem{Lonnblad:2012hz}
  L.~L\"onnblad,
  %``Fooling Around with the Sudakov Veto Algorithm,''
  Eur.\ Phys.\ J.\ C {\bf 73} (2013) no.3,  2350
  doi:10.1140/epjc/s10052-013-2350-9
  [arXiv:1211.7204 [hep-ph]].
  %%CITATION = doi:10.1140/epjc/s10052-013-2350-9;%%
  %3 citations counted in INSPIRE as of 15 Apr 2016

%\cite{Dulat:2015mca}
\bibitem{Dulat:2015mca}
  S.~Dulat {\it et al.},
  %``New parton distribution functions from a global analysis of quantum chromodynamics,''
  Phys.\ Rev.\ D {\bf 93} (2016) no.3,  033006
  doi:10.1103/PhysRevD.93.033006
  [arXiv:1506.07443 [hep-ph]].
  %%CITATION = doi:10.1103/PhysRevD.93.033006;%%
  %194 citations counted in INSPIRE as of 29 Aug 2016

%\cite{Schmidt:2015zda}
\bibitem{Schmidt:2015zda}
  C.~Schmidt, J.~Pumplin, D.~Stump and C.~P.~Yuan,
  %``CT14QED parton distribution functions from isolated photon production in deep inelastic scattering,''
  Phys.\ Rev.\ D {\bf 93} (2016) no.11,  114015
  doi:10.1103/PhysRevD.93.114015
  [arXiv:1509.02905 [hep-ph]].
  %%CITATION = doi:10.1103/PhysRevD.93.114015;%%
  %22 citations counted in INSPIRE as of 29 Aug 2016

%\cite{Bourhis:1997yu}
\bibitem{Bourhis:1997yu}
  L.~Bourhis, M.~Fontannaz and J.~P.~Guillet,
  %``Quarks and gluon fragmentation functions into photons,''
  Eur.\ Phys.\ J.\ C {\bf 2} (1998) 529
  doi:10.1007/s100520050158
  [hep-ph/9704447].
  %%CITATION = doi:10.1007/s100520050158;%%
  %216 citations counted in INSPIRE as of 30 Aug 2016

%\cite{Adare:2012vn}
\bibitem{Adare:2012vn}
  A.~Adare {\it et al.} [PHENIX Collaboration],
  %``Direct photon production in $d+$Au collisions at $\sqrt{s_{NN}}=200$ GeV,''
  Phys.\ Rev.\ C {\bf 87} (2013) 054907
  doi:10.1103/PhysRevC.87.054907
  [arXiv:1208.1234 [nucl-ex]].
  %%CITATION = doi:10.1103/PhysRevC.87.054907;%%
  %33 citations counted in INSPIRE as of 30 Aug 2016

%\cite{Adare:2012yt}
\bibitem{Adare:2012yt}
  A.~Adare {\it et al.} [PHENIX Collaboration],
  %``Direct-Photon Production in $p+p$ Collisions at $\sqrt{s}=200$ GeV at Midrapidity,''
  Phys.\ Rev.\ D {\bf 86} (2012) 072008
  doi:10.1103/PhysRevD.86.072008
  [arXiv:1205.5533 [hep-ex]].
  %%CITATION = doi:10.1103/PhysRevD.86.072008;%%
  %34 citations counted in INSPIRE as of 30 Aug 2016

%\cite{Gluck:1992zx}
\bibitem{Gluck:1992zx}
  M.~Gl\"uck, E.~Reya and A.~Vogt,
  %``Parton fragmentation into photons beyond the leading order,''
  Phys.\ Rev.\ D {\bf 48} (1993) 116
   Erratum: [Phys.\ Rev.\ D {\bf 51} (1995) 1427].
  doi:10.1103/PhysRevD.51.1427, 10.1103/PhysRevD.48.116
  %%CITATION = doi:10.1103/PhysRevD.51.1427, 10.1103/PhysRevD.48.116;%%
  %144 citations counted in INSPIRE as of 14 Sep 2016

%\cite{Klasen:1995xe}
\bibitem{Klasen:1995xe}
  M.~Klasen and G.~Kramer,
  %``Dijet cross-sections at o (alpha alpha-s**2) in photon - proton collisions,''
  Phys.\ Lett.\ B {\bf 366} (1996) 385
  doi:10.1016/0370-2693(95)01352-0
  [hep-ph/9508337].
  %%CITATION = doi:10.1016/0370-2693(95)01352-0;%%
  %88 citations counted in INSPIRE as of 13 Sep 2016

%\cite{Saimpert:2015oka}
\bibitem{Saimpert:2015oka}
  M.~Saimpert [ATLAS Collaboration],
  %``Measurement of Photon Production Cross Sections with the ATLAS Detector,''
  PoS DIS {\bf 2015} (2015) 155.
  %%CITATION = POSCI,DIS2015,155;%%

%\cite{KunnawalkamElayavalli:2016ttl}
\bibitem{KunnawalkamElayavalli:2016ttl}
  R.~Kunnawalkam Elayavalli and K.~C.~Zapp,
  %``Simulating V+jet processes in heavy ion collisions with JEWEL,''
  arXiv:1608.03099 [hep-ph].
  %%CITATION = ARXIV:1608.03099;%%

\end{thebibliography}

\end{document}